\documentclass[pr1,onecolumn,superscriptaddress,showpacs,amsmath,amssymb]{revtex4-1}
\usepackage{graphicx}
\usepackage{indentfirst}
\usepackage{psfrag}
\usepackage{epsfig}
\usepackage{amsmath}
\usepackage{amssymb}
\usepackage{bm}
\usepackage{color}
\usepackage[mathscr]{euscript}
\usepackage{mathrsfs}
\newcommand{\be}{\begin{eqnarray}}
\newcommand{\ee}{\end{eqnarray}}
\newcommand{\al}{{\it{et al.}}}
\newcommand{\hel}{\mathscr{ H}}
\newcommand{\leh}{\mathscr{ L}}
\newcommand{\flow} {\mathscr{ F}}
\newcommand{\perd}{\mathscr{ P}}
\newcommand{\qui}{\mathscr{ K}}
\newcommand{\s}{\mathscr{ S}}
\newcommand{\pp}{\mathscr{ PP}}

\newcommand{\bb}{\color{black}}

\begin{document}
\title{Active, reactive  and instantaneous optical forces on small particles in the time domain: Ultrafast  attosecond subcycle pulses }  
\author{Xiaohao Xu}
\affiliation{State Key Laboratory of Ultrafast Optical Science and Technology, Xi'an Institute of Optics and Precision Mechanics, Chinese Academy of Sciences, Xi'an 710119, China }
\author{F.J.  Valdivia-Valero}
\affiliation{Departamento de Física Aplicada,  Facultad de Ciencias del Mar y Ambientales, Universidad de Cádiz, Avenida de la Republica Saharahui s/n, Puerto Real, 11510 Cádiz, Spain.}
\author{M. Nieto-Vesperinas*}
\affiliation{Instituto de Ciencia de Materiales de Madrid, Consejo Superior de
Investigaciones Cient\'{i}ficas.\\
 Campus de Cantoblanco, Madrid 28049, Spain. }
\email{ Author to whom any correspondence should be addressed, mnieto@icmm.csic.es}
\pacs{or key words: radiation pressure, optical forces, reactive electromagnetic forces, time-dependent electrodynamical forces, dipolar nano-particles, evanescent waves, lateral forces, pulling forces, ultrafast optics, subcycle pulses.}
\begin{abstract}
Recently discovered reactive optical forces have nule time-average of their instantaneous values on monochromatic illumination, so that their detection suggests the use of ultrafast optics, specially in the  femto and attosecond domains. By using illumination with subcycle attosecond evanescent pulses, we report a theoretical study of  the time variations of  instantaneous forces and the behaviour of reactive forces versus  those active \bb  on small resonant particles that we consider dipolar.  We demonstrate how the structure of these pulses permit to obtain  three remarkable novel effects on electric dipoles; namely,  a lateral force, a pulling force against the canonical and Poynting momenta of the wavefield, and a  levitating effect  on the particle under repetition of the pulse. We expect that this study inaugurates a novel research in the area of optical manipulation. Future developments and experiments based on this theory should increase the insight and operation of the ultrafast dynamics of nanostructures. \bb
\end{abstract}
\maketitle


\section{Introduction}
The physics of optical forces, and the mechanical action of electromagnetic fields in general,  shows that there exist two quantities underneath their effect which we recently uncovered; namely, the reactive strength of canonical (or orbital) momentum and the reactive force \cite{LSA22}. Both of them act on real forces
\cite{jackson,griffiths,ptrs1,LSA17, qiuAOP,bliokh0}  in an analogous way as the   reactive power does in the  complex work in antenna radiation and electrical energy transmission \cite{guillemin,jackson,harrington,geyi,balanis,cpv}.

Reaching a minimum  in reactive power  has been a workhorse in RF antennas  and transmission line design. Its effects are well-known,  as a large reactive energy  involves  high ohmic loss and a decrease of operative performance. On the other hand, in the dynamics studies  in the nano-optics realm,  monochromatic ilumination has been generally addressed;  so that almost systematically time-harmonic wavefields have been considered. Then, subwavelength resolution from measurements of the mechanical action of  near-fields, developed in the last decades,  is not generally accompanied by a high temporal resolution. Hence, so far particle localization and manipulation  of nanostructures seem not helped by current high temporal resolution in nano-metrology.

In this work we combine the availability of advances in attosecond optics  with the discovery  of reactive  quantities associated to electrodynamic forces  acting on nano-objects which we shall model as dipolar spheres. We study the time variation of the instantaneous, {\it active} and reactive optical forces from a pulsed evanescent wave on a  small particle in which only its electric dipole is excited.  We expect that progress in this  subject may open an scenario of increasing understanding  of  ultrafast manipulation in light-matter interactions.

There is an essential role of the canonical (or orbital) momentum (OM), and also of the spin momentum (SM), in electromagnetic forces. So that the latter may be fully defined in terms of the real and imaginary parts of these momenta. The transfer of linear momentum to produce the optical force amounts to the storage of both OM and SM  in and around the object on which the electrodynamic force acts. 

As shown in \cite{LSA22} the {\it complex Lorentz force} and  the {\it complex Maxwell stress tensor} (CMST), of which its real part is the   {\it  active} \bb  Lorentz force (RLF) or active radiation pressure, (which   in the case of  monochromatic illumination \bb coincides with the time-averaged force),  contains an imaginary part which describes the reactive optical force (ILF) which is given in terms of   the {\it reactive strength of orbital momentum} (ROM) both  inside and outside the  object, as well as of  the {\it reactive spin momentum}. The latter amounts to the flow of this ROM through a surface that encloses the body, and is given by the flow of the imaginary part (IMST) of the complex Maxwell stress tensor (CMST). 

There is an antagonistic effect between the RLF and ILF forces; so that a large ILF and  storage of ROM, amounts to a loss of radiation force  (i.e. of RLF), and vice-versa.  Hence, the ILF acts as a predictor of the strength that we can expect of the radiation pressure RLF  on an object.

 We emphasize that in contrast with previous studies developed so far on pulsed forces, (see e.g. \cite{gong} and references therein), our RLF cannot be considered as a time-averaged force,  neither it is an instantaneous Lorentz force \cite{martin}, since the width of our ultrafast pulses, being subcycle ones, is smaller than their carrier period and hence, at difference with previous developments on pulsed forces, they cannot be considered as effectively quasi-monochromatic in  their carrier frequency. This introduces a clear distinction in the active force, RLF, because although it is is defined in terms of the analytic signals associated to the fields \cite{LSA22} and it is averaged over a reactive time scale $s$ \cite{LSA22, kaiser}, in the zero limit, $s \rightarrow 0$, of this scale it cannot be associated with the local value of the instantaneous Lorentz force, neither with its average over one period of its carrier frequency. This is due to the fact that no such connection exists between the  CMST theorem for our pulses and the CMST theorem for quasi-monochromatic (i.e. time-harmonic) fields. 

 As a consequence of this study, we demonstrate   that the structure of these pulses allow a great flexibility in  designing  the resulting forces that they exert on matter. In this way we show  three novel dynamical effects from  a subcycle pulsed evanescent wave  on an electric dipole sphere resting on a total internal reflection (TIR) interface. These are: first,  a lateral force, pushing or pulling according to the choice of the  carrier phase and time-shift of the field, second, a pulling force acting against the canonical and Poynting momenta of the evanescent pulse. It should be reamarked that so far both effects had only been found on magnetoelectric dipolar particles \cite{cpv, bliokh1}, thus being so far  considered impossible on purely electric dipoles. Third, there exists  a remarkable  levitating  action of the pulsed evanescent wave,  thus pushing the particle away from the interface under repetition of the pulse. \bb

\section{Time-dependent fields: The conservation of complex momentum  and the active real and reactive forces}
  Being $\bm{\mathcal  E}({\bf r},t)$  and $\bm{\mathcal  B}({\bf r},t) $  the analytic signals associated to the electric and magnetic fields, (cf. Eq. (1) of \cite{LSA22}) in the limit of reactive time scale $s \rightarrow 0$, and    $\bm{\mathcal G}({\bf r},t)=(1/c^2)\bm{\mathcal  S}({\bf r},t)=(1/8\pi c) \bm{\mathcal  E}({\bf r},t)   \times \bm{\mathcal  B}^*({\bf r},t) $ the complex Poynting momentum density, the conservation equation of the  complex mechanical momentum $\bm{\mathcal {P}_{mech}}$ on a body reads \cite{LSA22}:
\be\partial_t[\mathcal {P}_{mech\, i}+\int_{V}d^3 r \, {\mathcal G}_i^*]= \int_{\partial V}d^2 r \, {\mathcal T}_{ij} 
n_j+i\omega\int_{V}d^3 r\, [\bm{\mathcal P}_{m}^O - \bm{\mathcal P}_{e}^O]_i\, . \,\,\,  \label{tfor36}
\ee
Where $V$ is a volume surrounding the object, $\omega$ is a carrier frequency and  the {\it  complex Maxwell stress tensor} (CMST) reads
\be
{\mathcal T}_{ij}({\bf r},t)=\frac{1}{8\pi}[{\mathcal E}_i  {\mathcal E}_j^* + {\mathcal B}_i ^*{\mathcal B}_j -\frac{1}{2}\delta_{ij}(|\bm{\mathcal E}|^2 + |\bm{\mathcal B}|^2 )], \,\,\,\,\, (i,j=1,2,3).\,\, \label{tfor34}
\ee
And with the appearence in  (\ref{tfor36}) of the canonical momentum densities:

\be  
(\bm{ P}_{e}^O)_i({\bf r},t)  =\frac{1}{8\pi \omega }\mbox{Im}  [{\mathcal E}_j^* \partial_i {\mathcal E}_j], \,\,\,\,\,\,\,\,\,
(\bm{\mathcal P}_{m}^O)_i({\bf r},t) = \frac{1}{8\pi \omega }\mbox{Im}[ {\mathcal B}_j^* \partial_i {\mathcal B}_j ],).\,\,\, \label{tfor35}
\ee

The real and imaginary parts of (\ref{tfor34}) yield
\be 
{\mathcal F}_i^R({\bf r},t)\equiv\partial_t { P}_{mech \, i}^R=-\int_{V}d^3 r \, \partial_t {\mathcal G}_i^{R}+ \int_{\partial V}d^2 r \,  {\mathcal T}_{ij}^R n_j\,, \,\,\,  \label{tfor37a}\, \,\,\, \,\,\, \,\,\\
 {\mathcal T}_{ij}^R=\frac{1}{8\pi}\Re[ {\mathcal E}_i   {\mathcal E}_j^* +  {\mathcal B}_i ^* {\mathcal B}_j] -\frac{1}{2}\delta_{ij}(| \bm{\mathcal E}|^2 + | \bm{\mathcal B}|^2 ); \nonumber
\ee
and
\be 
 {\mathcal F}_i^I({\bf r},t)\equiv\partial_t { P}_{mech\, i}^I=\int_{V}d^3 r \,\partial_t {\mathcal G}_i^{I}
+ \int_{\partial V}d^2 r \, {\mathcal T}_{ij}^I n_j+\int_{V}d^3 r\, [ \bm{\mathcal P}_{m}^O-  
\bm{\mathcal P}_{e}^O]_i\,. \,\,\, \,\, \,\, \,\,\,   \label{tfor37b}\\
 \bm{\mathcal T}_{ij}^I=\frac{1}{8\pi}\Im[ {\mathcal E}_i   {\mathcal E}_j^* +  {\mathcal B}_i ^* 
{\mathcal B}_j].\, \,\,\, \,\,\, \,\,\, \,\,\, \,\,\, \,\,\, \,\,\, \,\,\, \,\,\, \,\,\label{tfor37c}
\ee
All quantities in the above equations are functions of ${\bf r}$ and $t$. While (\ref{tfor37a}) describes how the increase of  momentum in the body, i.e. mechanical plus field (i.e. Poynting) momentum, equals the momentum brought in by the field  due to its incoming  flow,  Eq.  (\ref{tfor37b}) represents the change in the body  of an instantaneous  mechanical momentum minus the transfer of reactive (i.e. imaginary)  Poynting  momentum of the field; and this   equals the flow of this reactive momentum into the body volume accompanied by the accretion of   {\it reactive strength of canonical momentum}:  $\int_{V}d^3 r\, [\bm{\mathcal P}_{m}^O - \bm{\mathcal P}_{e}^O]$,  \cite{LSA22}.

 Thus the flow of reactive  momentum into the object volume plus the increase of reactive canonical (i.e. orbital) momentum of the field yields an  instantaneous mechanical  momentum in the body,  i.e. the source term of the conservation law, given by the ILF;  $\int_V Im\{\rho({\bf r},t)\,\bm{\mathcal E}({\bf r},t)+\frac{1}{c}\bm{\mathcal J}^*({\bf r},t)\times\bm{\mathcal B}({\bf r},t)\}$, ($\rho$ and $\bm{\mathcal J}$ being the charge and current densities in the body). Hence Eq.  (\ref{tfor37b}) constitutes the conservation law for the reactive field momentum.  For these reasons,  $\bm{\mathcal F}^I$ has been called the {\it reactive force} on the object  \cite{LSA22}, in contrast with the radiation pressure or {\it active real} {\it force} $\bm{\mathcal F}^R$. 

We shall see that the latter force  $\bm{\mathcal F}^R$ is part of an instantaneous force.

\section{The instantaneous, active and reactive forces on a dipolar particle}
The instantaneous Lorentz  force  exerted by a generic field of electric and magnetic vectors $\mathbb{E}=\Re[{\bf E}({\bf r}, t)]$ and $\mathbb{B}=\Re[{\bf B}({\bf r}, t)]$ on a wide-sense dipolar particle \cite{nieto1,torquePRA,patrickOL 2000} of dipolar moment $\mathbb{P}=\Re[{\bf p}({\bf r}, t)]$, is:  $\partial_t { \bf P}_{mech}\equiv {\bf F}({\bf r}, t)=[\mathbb{ P}\cdot \nabla]\mathbb{E}+(1/c)[\partial_t\mathbb{P}\times\mathbb{B}]$. Namely, 
\be
\partial_t { \bf P}_{mech}\equiv {\bf F}({\bf r}, t)=\frac{1}{2}\{\Re[{\bf p}^*({\bf r}, t) 
\cdot\nabla){\bf E}({\bf r}, t)]+\frac{1}{c}\Re[\partial_t {\bf p}^*({\bf r}, t) \times {\bf B}({\bf r}, t)] \nonumber \\
+\Re[{\bf p}({\bf r}, t) 
\cdot\nabla){\bf E}({\bf r}, t)]+\frac{1}{c}\Re[\partial_t {\bf p}({\bf r}, t) \times {\bf B}({\bf r}, t)]\}; \label{fdip1r}
\ee
the asterisk denoting complex conjugate. Since in terms of the particle polarizability $\alpha_e(t)$ the induced electric dipole moment holds: 
${\bf p}({\bf r}, t)=\alpha_e(t)  *  {\bf E}({\bf r}, t)$, this instantaneous force,  Eq. (\ref{fdip1r}),  is straightforwardly transformed by
 using $\nabla\times {\bf E} =-(1/c)\partial_t {\bf B}$ and the identity for solenoidal fields $
{\bf a}$:  $ {\bf a}^*\times(\nabla\times {\bf a})=a_j^* \partial_i a_j - a_j^* \partial_j a_i$ 
and  $ {\bf a}\times(\nabla\times {\bf a})=a_j\partial_i a_j - a_j \partial_j a_i$, into
\be
\partial_t { P}_i^{mech}\equiv {F}_i({\bf r}, t)=\frac{1}{2}\{\Re[{ p}_j^*({\bf r}, t)\partial_i{E}_j({\bf r}, t)]+\frac{1}{c}\Re\{\partial_t [{\bf p}^*({\bf r}, t) \times {\bf B}({\bf r}, t)]_i \} \nonumber \\
+\Re[{ p}_j({\bf r}, t)\partial_i{E}_j({\bf r}, t)]+\frac{1}{c}\Re\{\partial_t [{\bf p}({\bf r}, t) \times {\bf B}({\bf r}, t)]_i\}\}
, \,\,(i,j=1,2,3).\, \,\,\, \,\,\, \,\,\,\label{fdip2r}
\ee
Like in the case of the instantaneous complex Poynting theorem \cite{cpv,kaiser} and the instantaneous Maxwell stress tensor \cite{LSA22}, the first two terms of (\ref{fdip2r}) constitute  what we denominate the {\it active force},  ${F}_i^R({\bf r}, t)$,  or {\it time-dependent counterpart} of the well-known time- averaged Lorentz force exerted by time-harmonic -i.e. monochromatic- electromagnetic fields   \cite{ptrs1,nieto1,patrickOL 2000}; viz.,
\be
{F}_i^R({\bf r}, t)=\frac{1}{2}\{\Re[{p}_j^*({\bf r}, t)\partial_i{E}_j({\bf r}, t)]+\frac{1}{c}\partial_t \,[\Re\{{\bf p^*}({\bf r}, t) \times {\bf B}({\bf r}, t)\}]_i\}.\,\,\,\, \,\,\, \,\,\,\,\,\, \,\,\, \,\,\, \label{fdip3a}
\ee
Since ${\bf p}({\bf r},t)=\alpha(t)*{\bf E}({\bf r},t)$, (the symbol $*$  denotes convolution in $t$ and $\alpha(t)$ is the particle  complex polarizability), the active Lorentz force, Eq. (\ref{fdip3a}), describes the evolution of the gradient, orbital momentum, and the real and imaginary Poynting momentum \cite{cpv,nieto1,xu1} .

\begin{figure} [htbp]
\centerline{\includegraphics[width=1.0\columnwidth]{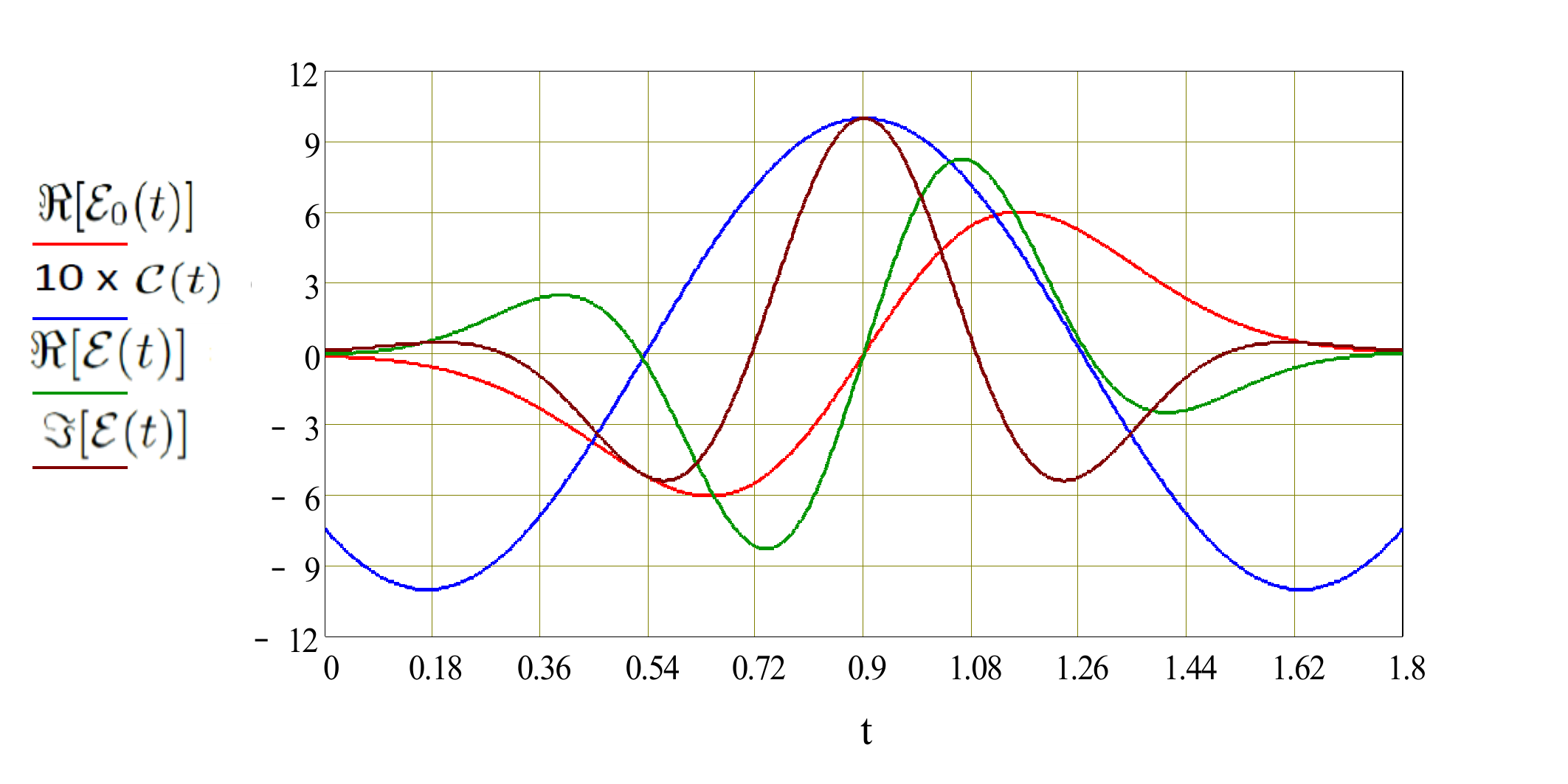}}
\caption{The pulse $\mathcal{E}(t)=\mathcal{E}_0(t)\exp(-i\omega_0 t)$ and the carrier c-wave, $x=0$, $\phi_0=0$. Real part of envelope $\Re[\mathcal{E}_0(t)]$, $\mathcal {C} (t)=A_0\Re[\exp(-i\omega_0 t)]$,   $\Re[\mathcal{E}(t)]=\Re [\mathcal{E}_0(t)\exp(-i\omega_0 t)]$,  $\Im[\mathcal{E}(t)]=\Im[\mathcal{E}_0(t)\exp(-i\omega_0 t) ]$. ($t$ in $fs$. The units of the ordinate are $10^{-9}\times10^{-3/2} J^{1/2}/ nm^{3/2}.$. See the magnitude and units of this incident pulse in the description of Fig. 2}  
\end{figure}

Instantaneous optical forces are relevant when one enters in the realm of attosecond optics \cite{corkum,ivanov,chang,chang1,jopt}.
Let the wave incident  on the dipolar particle be a pulse which we write in the form of a focused field whose main direction of propagation is along $OX$. We address wavefields for which one may write
\be
  {\bf E}({\bf r},t)   =  {\mathcal E}(t-\frac{x}{c})  {\bf E}_0({\bf r}) = {\mathcal E}_0(t-\frac{x}{c}){\bf E}_0({\bf r}) \exp[-i\omega_0( t-\frac{x}{c})]        , \,\,\,\\ \nonumber  
 {\bf B}({\bf r},t) =  {\mathcal E}(t-\frac{x}{c})  {\bf B}_0({\bf r}) ={\mathcal E}_0(t-\frac{x}{c}){\bf B}_0({\bf r}) \exp[-i\omega_0 ( t-\frac{x}{c})] \, ; \,\,\,\,\,\, \label{fdip3bis}
\ee
where $\omega_0$ is the carrier c-frequency. 
We thus consider subcycle pulses with a Gaussian time-dependent envelope  of half-width $\tau$, being  written as \cite{becker,kundu,astapenko}:
\be
{\mathcal E}(t-\frac{x}{c})={\mathcal E}_0(t-\frac{x}{c})\exp[-i\omega_0 ( t-\frac{x}{c})]  \nonumber \\
=iA_0 \frac{(1-i\frac{2(t-\frac{x}{c})}{\omega_0 \tau^2})^2+\frac{2}{\omega_0^2 \tau^2}}{1+\frac{2}{\omega_0^2 \tau^2}} \exp(-\frac{(t-\frac{x}{c})^2}{\tau^2})\exp[-i\omega_0 ( t-\frac{x}{c})-i\phi_0] . \label{fdip4bis}
\ee
Whose  Fourier spectrum is
\be
{\mathcal E} (\omega)= \frac{\sqrt{\pi}}{2}A_0 \tau \frac{i\omega^2\tau^2}{2+\omega_0^2 \tau^2}\exp[-\frac{\tau^2(\omega-\omega_0)^2}{4}-i\phi_0].
\, \,\,\,\,\, \,\,\,\,\, \,\, \label{fdip5bis}
\ee
$\phi_0$ being a carrier phase and $\tau<T_0=2\pi/\omega_0$.  
 Notice that $\int_{\infty}^{\infty}dt {\bf E}({\bf r},t)=\int_{\infty}^{\infty}dt {\bf B}({\bf r},t)={\bf 0}$, as it should \cite{kundu}. \bb

Since  in  Eqs. (\ref{fdip3a})-(\ref{fdip3b}) we are dealing with analytic signals in $t$, continued into $z=t+is$, which are  associated to the real fields \cite{LSA22}, stemming from the causality of the interaction, the spectrum given by Eq. (\ref{fdip5bis}) should vanish for $\omega<0$ \cite{LSA22}, therefore in what follows we drop the term with the factor $\exp[-\frac{(\omega+\omega_0)^2}{4}+i\phi_0]$.

As mentioned above,   the electric dipole moment ${\bf p}({\bf r}, t)$ holds:
\be 
{\bf p}^*({\bf r}, t)=\alpha_e^*(t)*{\bf E}^*({\bf r}, t) \, ; \label{fdip6}
\ee
 Since $\alpha_e^*(t)*{\bf E}^*({\bf r}, t)=\int_{-\infty}^{\infty} d\nu\exp(i2\pi\nu t)(\nu)\alpha^*(\nu){\bf E}^*({\bf r}, \nu)$,  one has  in frequency space, 

:
\be 
{\bf p}^*({\bf r}, \nu)=\alpha_e^*(\nu){\bf E}^*({\bf r}, \nu) \, ;  \,\,\, (\nu=\omega/2\pi).\,\,\,\label{fdip7}
\ee
Where  ${\bf E}^*({\bf r}, \nu)$, as well as ${\bf B}^*({\bf r}, \nu)$, are given in terms of  ${\mathcal E} (\nu)$, cf. Eq. (\ref{fdip5bis}), as
\be
{\bf E}^*({\bf r}, \nu)={\mathcal E}^* (\nu){\bf E}_0^*({\bf r}) , \,\,\,\,\,\,\,\,
{\bf B}^*({\bf r}, \nu)={\mathcal E}^* (\nu){\bf B}_0^*({\bf r})
 \label{fdip8}
\ee
I.e.
\be 
{\bf p}^*({\bf r}, \nu)=\alpha_e^*(\nu){\mathcal E}^* (\nu){\bf E}_0^*({\bf r})={\mathcal P}^*(\nu) {\bf E}_0^*({\bf r}). \,\,\,\,\,\,\,\, {\mathcal P}^*(\nu)=\alpha_e^*(\nu){\mathcal E}^*(\nu)
. \,\,\,\,\label{fdip9}
\ee
And
\be 
{\bf p}^*({\bf r}, t)=\alpha_e^*(t)*{\mathcal E}^*(t) {\bf E}_0^*({\bf r})={\mathcal P}^*(t) {\bf E}_0^*({\bf r}). \,\,\,\,\,\,\,\, {\mathcal P}^*(t)=\alpha_e^*(t)*{\mathcal E}^*(t).\,\,\,\,\,\,\, \,\,\,\,\,\,\,\,\,\,\,\label{fdip10}
\ee
We see  that for the time-dependent fields addressed,  there is a separation of $t$ and ${\bf r}$-dependent  factors which also holds in each term of the forces. This and the above Fourier analysis show that for each $\nu$-component we may proceed like for a monochromatic field as in \cite{LSA22}. Therefore Eq. (\ref{fdip3a})  may be considered in the $t$-domain as the real  part of the quantity:
\be 
\frac{1}{2}\{{p}_j^*({\bf r}, t)\partial_i{E}_j({\bf r}, t)+\frac{1}{c}\partial_t [{\bf p^*}
({\bf r}, t) \times {\bf B}({\bf r}, t)]_i\}\\
=\frac{1}{2}\{[{\mathcal P}^*(t)\mathcal{E}(t)][{ E}_{0\, j}^*({\bf r})\partial_i {E}_{0\, j}({\bf r})]+\frac{1}{c}\partial_t[{\mathcal P}^*(t)\mathcal{E}(t)][{\bf E}_0^*({\bf r})\times {\bf B}_0({\bf r})]_i\},  \,\,\,\,\,\,\,\,\,\,\,\,\,\,\,\,\,\,\label{fdip10aa}   \\
 (i,j=1,2,3).\,\,\,\,\,\,\,\,\, \nonumber 
\ee
Hence we now we introduce the {\it reactive Lorentz force} as the imaginary part of this quantity.  Namely:
\be
{ F}_i^I({\bf r}, t)=\frac{1}{2}\{\Im[{p}_j^*({\bf r}, t)\partial_i{E}_j({\bf r}, t)]+\frac{1}{2}\partial_t \,[\Im\{{\bf p^*}({\bf r}, t) \times {\bf B}({\bf r}, t)\}]_i\} , \,\,\,\, \,\,\, \,\,\,\,\,\, \,\,\, \,\,\, \label{fdip3b}
\ee
which should correspond to Eq.(\ref{tfor37b}) and  also shows the appearence of the gradient, orbital momentum, and the real and imaginary Poynting momentum.

Given the unexplored territory of these forces  ${ \bf F}^R$ and ${ \bf F}^I$, all we can say by now is that  they constitute generalizations for time-dependent wavefields of period-average and imaginary Lorentz forces for monochromatic  fields. More theoretical and experimental research is necessary to fully  identify their influence in the Lorentz force. 

So we get for the time-domain active and reactive forces:
\be
{ F}_i^R({\bf r}, t)=\frac{1}{2}\{\Re[{\mathcal P}^*(t)\mathcal{E}(t)]\Re[{ E}_{0\, j}^*({\bf r})\partial_i {E}_{0\, j}({\bf r})]-\Im[{\mathcal P}^*(t)\mathcal{E}(t)]\Im[{ E}_{0\, j}^*({\bf r})\partial_i { E}_{0\, j}({\bf r})] \nonumber \\
+\frac{1}{c}\Re[\partial_t({\mathcal P}^*(t)\mathcal{E}(t))]\Re[{\bf E}_0({\bf r})\times {\bf B}_0^*({\bf r})]_i +\frac{1}{c}\Im[\partial_t({\mathcal P}^*(t)\mathcal{E}(t))]\Im[{\bf E}_0({\bf r})\times {\bf B}_0^*({\bf r})]_i\} .
\,\,\,\,\,\,\,\,\,\,\,\,\, \label{fdip10_1}
\ee
\be
{ F}_i^I({\bf r}, t)=\frac{1}{2}\{\Re[{\mathcal P}^*(t)\mathcal{E}(t)]\Im[{E}_{0\, j}^*({\bf r})\partial_i { E}_{0\, j}({\bf r})]+\Im[{\mathcal P}^*(t)\mathcal{E}(t)]\Re[{ E}_{0\, j}^*({\bf r})\partial_i { E}_{0\, j}({\bf r})] \nonumber 
\\  -\frac{1}{c}\Re[\partial_t({\mathcal P}^*(t)\mathcal{E}(t))]\Im[{\bf E}_0({\bf r})\times {\bf B}_0^*({\bf r})]_i +\frac{1}{c}\Im[\partial_t({\mathcal P}^*(t)\mathcal{E}(t))]\Re[{\bf E}_0({\bf r})\times {\bf B}_0^*({\bf r})]_i\} , \,\,\,\,\,\,\,\,\,\,\,\, \label{fdip10_2}
\ee
respectively. The first and second terms of (\ref{fdip10_1}) contain the  gradient component proportional to $\partial_i |{\bf E}_0({\bf r})|^2$ and the orbital momentum of the spatial part of the fields, given by $\Im[{ E}_{0\, j}^*({\bf r})\partial_i {E}_{0\, j}({\bf r})]$, which comprises the spin-curl  and the Poynting momentum foces. Reciprocally occurs in Eq. (\ref{fdip10_2}).  These are  well-known  components from monochromatic  continuous waves on  time-averaging. In fact, in this latter case  one would have: ${\mathcal P}^*(t)=\alpha_e^*(\nu) \exp(i\nu_0 t)$ and  $\mathcal{E}(t)=\exp(-i\nu_0 t)$; therefore  the time derivative terms of  (\ref{fdip10_1}) and (\ref{fdip10_2}) would vanish and only the second term of  (\ref{fdip10_2}) would contribute to  ${\bf F}^I$; then Eqs.(\ref{fdip10_1}) and (\ref{fdip10_2}) would reduce to those well-known active (or time-averaged) real and reactive  electromagnetic forces on an electric dipole, respectively,  \cite{LSA22,patrickOL 2000}. 

However, for the pulse, the contribution of the complex momentum $\frac{1}{2c}[{\mathcal P}^*(t)\mathcal{E}(t)][{\bf E}_0^*({\bf r})\times {\bf B}_0({\bf r})]$,  described by the second term of (\ref{fdip10aa}), makes both the real and the imaginary Poynting momenta to come into play, as seen in the last two terms of  (\ref{fdip10_1}),  with an analogous contribution  in  the reactive force (\ref{fdip10_2}).

Analogously, the instantaneous Lorentz force, Eq. (\ref{fdip2r}), becomes:
\be
{F}_i({\bf r}, t)={ F}_i^R({\bf r}, t)+{ F'}_i({\bf r}, t); \label{fdip10_1r}
\ee
where
\be
{ F'}_i({\bf r}, t)=\frac{1}{2}\{\Re[{\mathcal P}(t)\mathcal{E}(t)]\Re[{ E}_{0\, j}({\bf r})\partial_i {E}_{0\, j}({\bf r})]-\Im[{\mathcal P}(t)\mathcal{E}(t)]\Im[{ E}_{0\, j}({\bf r})\partial_i { E}_{0\, j}({\bf r})] \nonumber \\
+\frac{1}{c}\Re[\partial_t({\mathcal P}(t)\mathcal{E}(t))]\Re[{\bf E}_0({\bf r})\times {\bf B}_0({\bf r})]_i -\frac{1}{c}\Im[\partial_t({\mathcal P}(t)\mathcal{E}(t))]\Im[{\bf E}_0({\bf r})\times {\bf B}_0({\bf r})]_i\} .
\,\,\,\,\,\,\,\,\,\,\,\,\, \label{fdip10_1prim}
\ee
stems from the third and fourth terms of (\ref{fdip2r}).
The separation of factors of $t$ and ${\bf r}$  in Eqs.  (\ref{fdip10aa})-(\ref{fdip10_1r}) indicates that, given $\mathcal{E}(t-x/c)$, one has to find the dipole $t$-factor ${\mathcal P}(t)$ (cf. Eq.(\ref{fdip10})) in order to work out these equations, since these $t$-dependent functions rule the time-resolved forces exerted by the pulse.   Hence,  we now concentrate in the temporal factors:   ${\mathcal P}(t)\mathcal{E}(t)$, $(1/c)[\partial_t({\mathcal P}(t)\mathcal{E}(t))]$, $(1/2)[{\mathcal P}^*(t)\mathcal{E}(t)]$ and $(1/2c)[\partial_t({\mathcal P}^*(t)\mathcal{E}(t))]$.

It is known \cite{cpv,bliokh1} that, for instance,  an evanescent wave may have a complex Poynting momentum  with a transversal $y$-component.   This fact has led to studies \cite{cpv,bliokh1,AOP} on the transversal time-averaged force exerted by monochromatic evanescent waves on magnetodielectric particles, stemming from to the second order interaction on interference between the  electric and magnetic dipoles induced  in the scattering particle.  In turn, the last two terms of  (\ref{fdip10_1})-(\ref{fdip10_2}) state that an evanescent pulse may exert an instantaneous first-order transversal force even on a purely electric dipolar particle, (the same could equally be derived on a purely magnetic dipolar particle). This is  seen in detail below.

 \begin{figure} [htbp]
\centerline{\includegraphics[width=1.0\columnwidth]{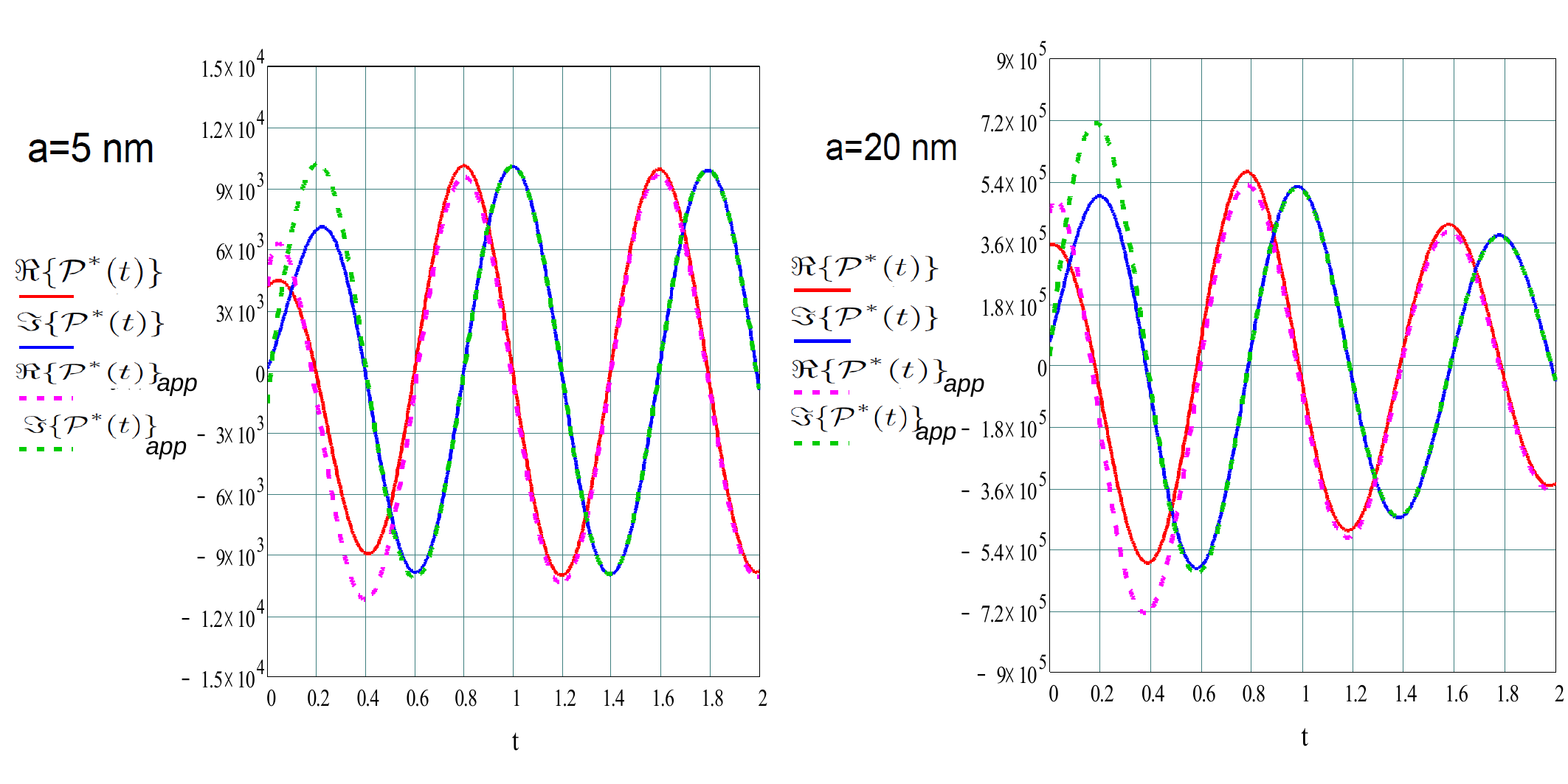}}
\caption{ ${\Re\{\mathcal P}^*(t)\}$ and ${\Im\{\mathcal P}^*(t)\}$.  $t$ in $fs$.   The units of the ordinate are
$10^{-9}\times10^{-3/2} J^{1/2}\times nm^{3/2}$. Full lines: Exact values. Broken lines and subindex $app$: Approximation Eq. (\ref{fdip11}). Left:  Ag particle with radius $a=5 nm$. Right:  $a=20 nm$. These particles have the resonant frequency: $\nu_{r}=1261 THz$ and $\nu_{r}=1259 THz$, respectively. Thus the oscillation period is $1/\nu_{r}=0.793 fs$ and  $1/\nu_{r}=0.794 fs$, respectively.}
\end{figure}

\begin{figure} [htbp]
\centerline{\includegraphics[width=1.0\columnwidth]{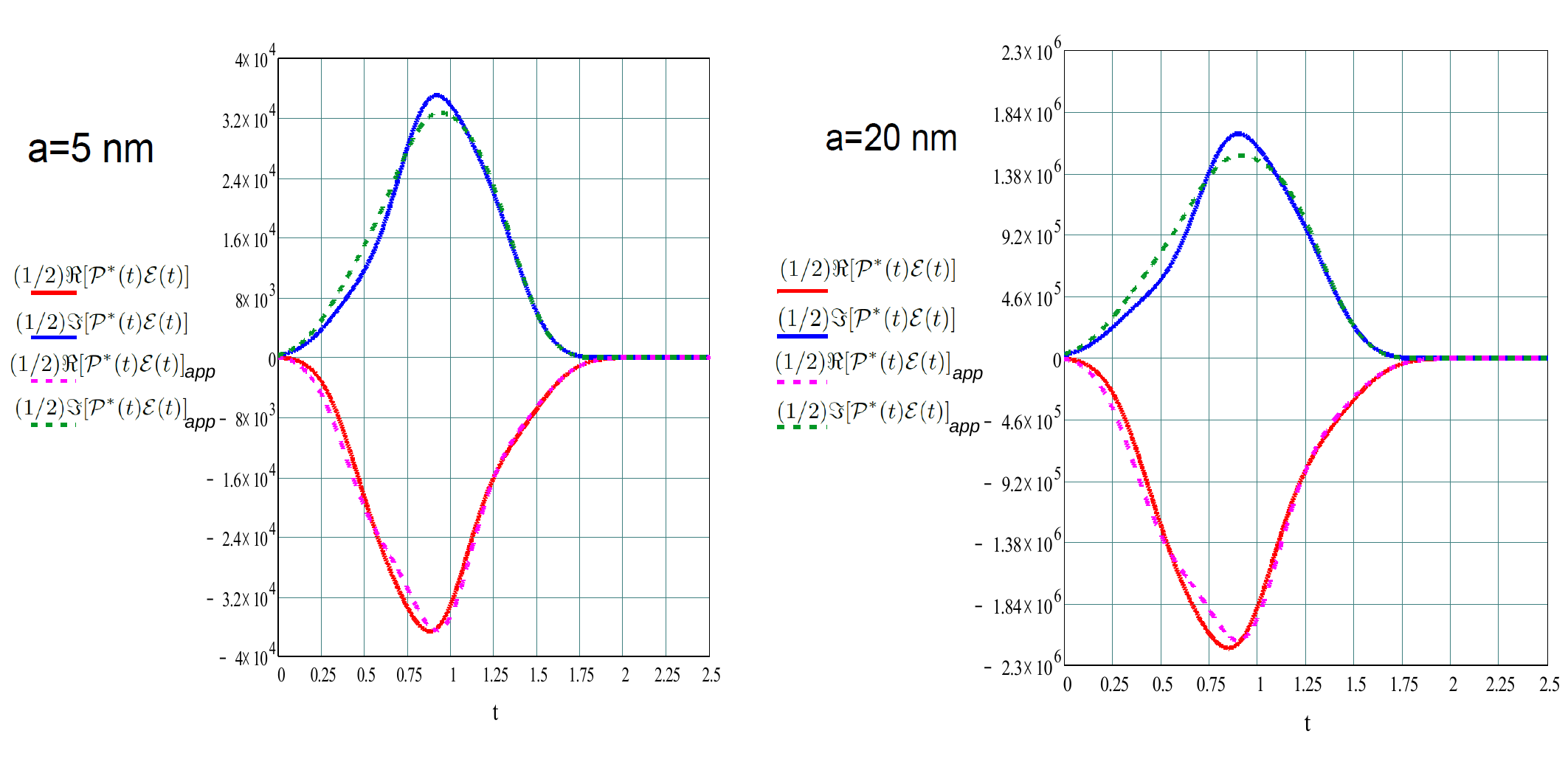}}
\caption{ $(1/2)\Re[{\mathcal P}^*(t)\mathcal{E}(t)]$ and 
$(1/2)\Im[{\mathcal P}^*(t)\mathcal{E}(t)]$.  $t$ in $fs$.  The units of the ordinate are $10^{-21} J$.
Full lines: Exact values. Broken lines and subindex $app$: given by the
approximation Eq. (\ref{fdip11}). Left:  Ag particle with radius $a=5 nm$. Right:   $a=20 nm$.}
\end{figure} 

\begin{figure} [htbp]
\centerline{\includegraphics[width=1.0\columnwidth]{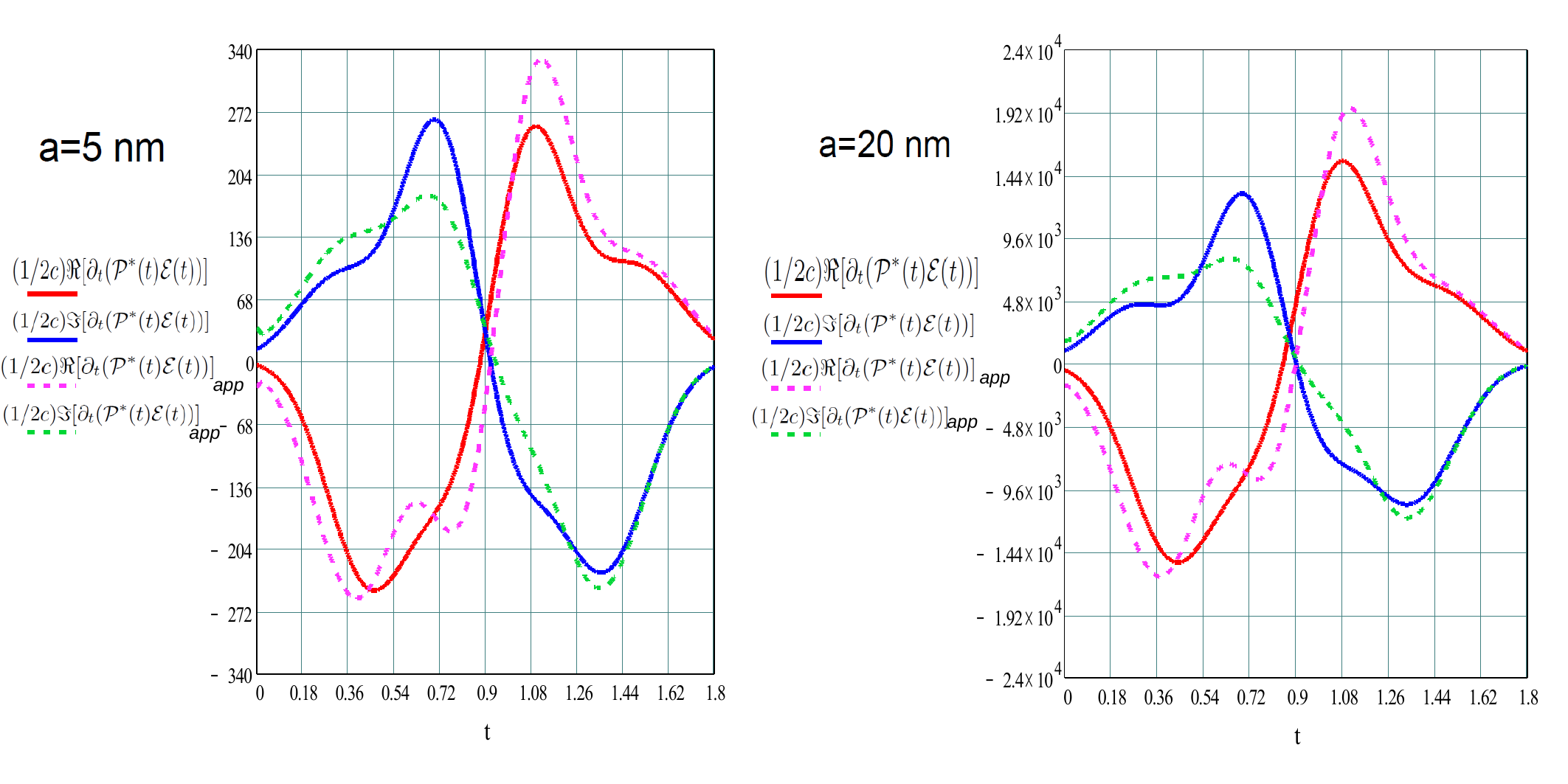}}
\caption{ $(1/2c)\Re[\partial_t({\mathcal P}^*(t)\mathcal{E}(t))]$ and 
$(1/2c)\Im[\partial_t({\mathcal P}^*(t)\mathcal{E}(t))]$.  $t$ in $fs$.  The units of the ordinate are $pN$. Full lines: Exact values. Broken lines and subindex $app$: given by the
approximation Eq. (\ref{fdip11}). Left:  Ag particle with $a=5 nm$. Right:   $a=20 nm$.}
\end{figure}

\begin{figure}
\centerline{\includegraphics[width=1.0\columnwidth]{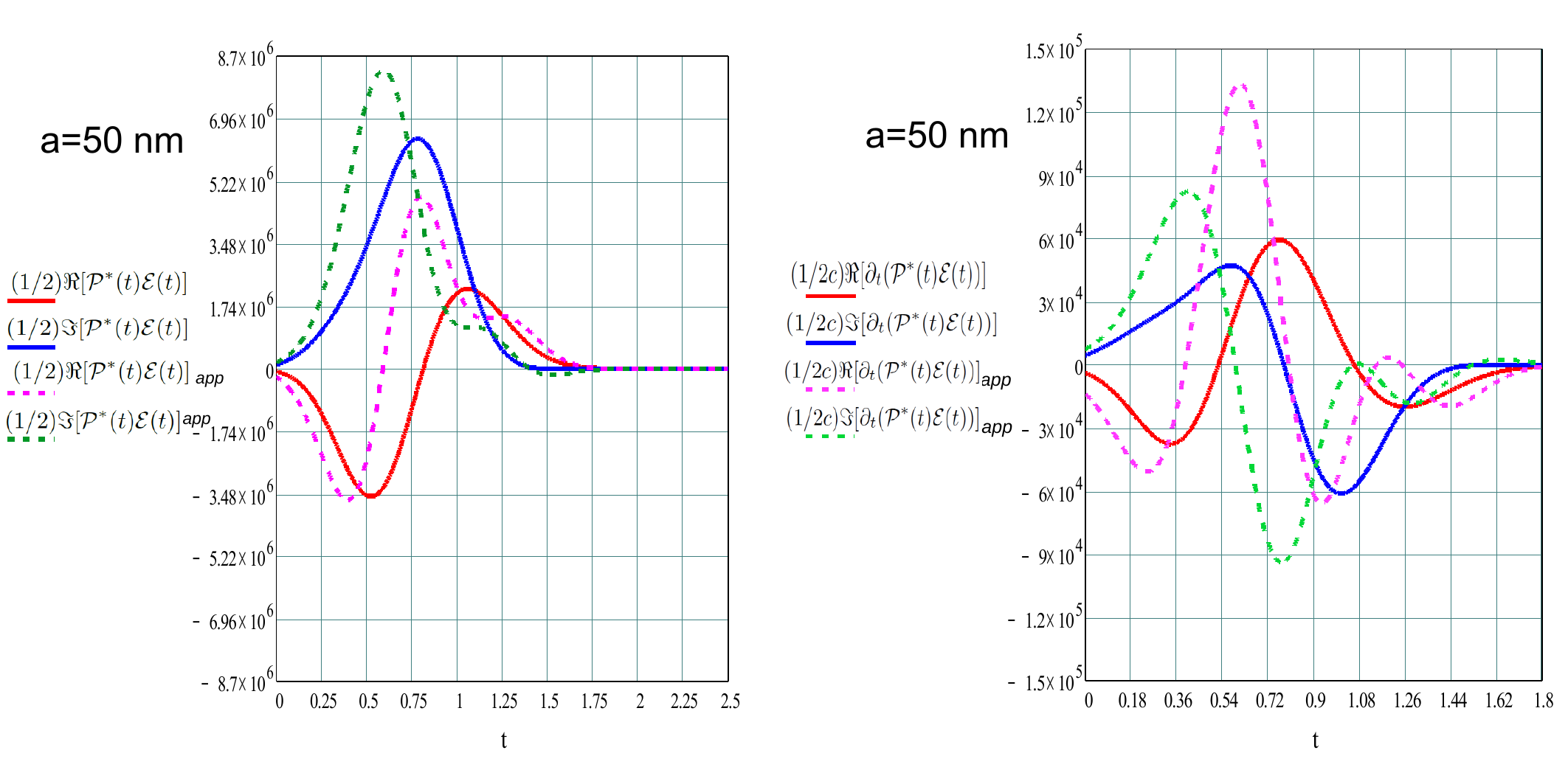}}
\caption{ Departure of the approximation Eq.(\ref{fdip11}) in  $(1/2){\mathcal P}^*(t)\mathcal{E}(t)$  (left),  and $(1/2c)\partial_t({\mathcal P}^*(t)\mathcal{E}(t))$ (right)  for an Ag sphere with $a=50 nm$. Solid lines are exact values; broken lines correspond to the approximation Eq. (\ref{fdip11}). $t$ in $fs$.  \bb }
\end{figure}

\begin{figure} [htbp]
\centerline{\includegraphics[width=1.0\columnwidth]{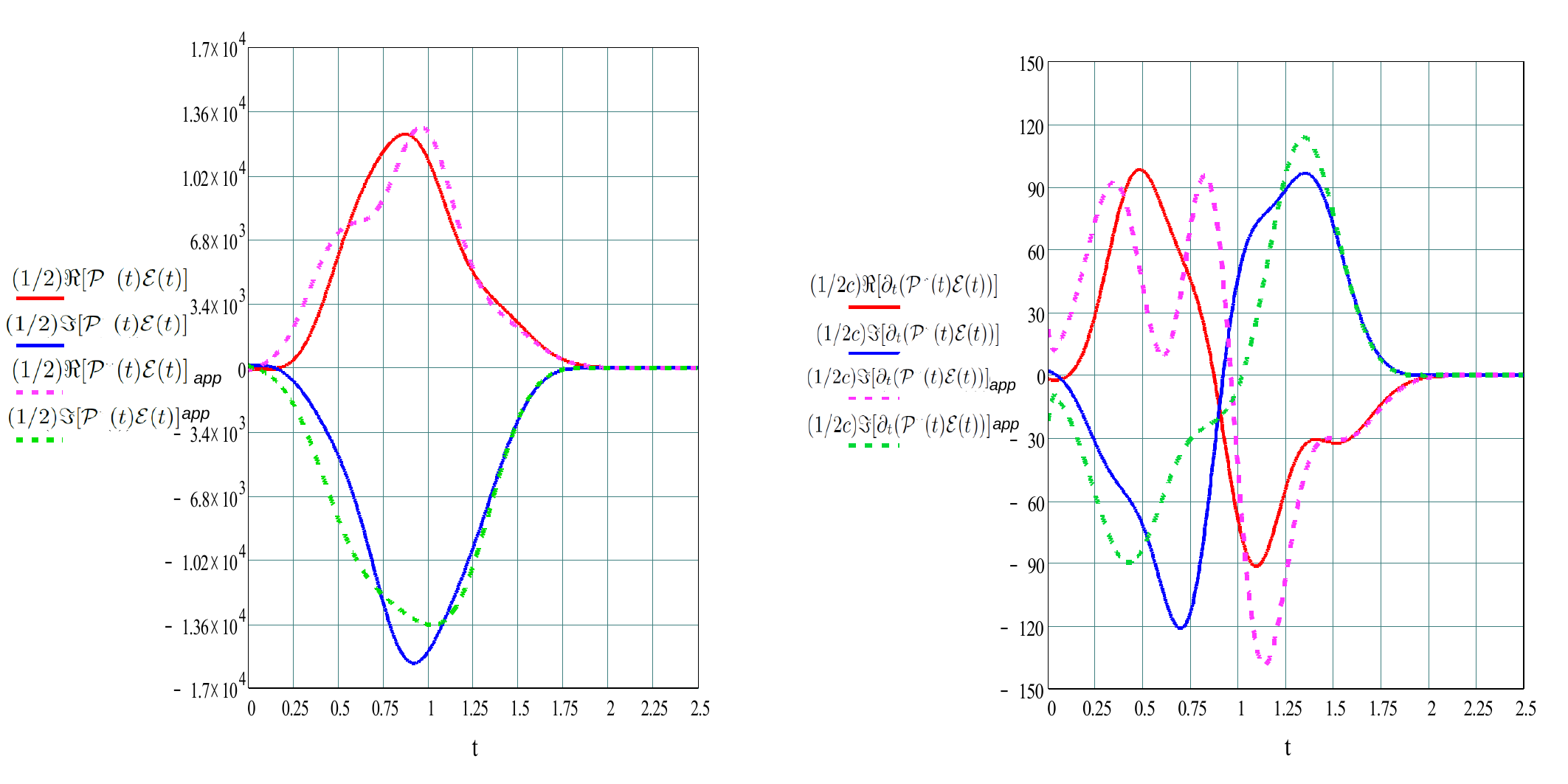}}
\caption{ Ag particle with radius $a=5nm$.  Left: $(1/2)\Re[{\mathcal P}(t)\mathcal{E}(t)]$ and  $(1/2)\Im[{\mathcal P}(t)\mathcal{E}(t)]$.  $t$ in $fs$.  The units of the ordinate are  $10^{-21} J$ . Right: $(1/2c)\Re[\partial_t({\mathcal P}(t)\mathcal{E}(t))]$ and $(1/2c) \Im[\partial_t({\mathcal P}(t)\mathcal{E}(t))]$.  $t$ in $fs$.  The units of the ordinate are  $pN$. Full lines: Exact values. Broken lines and subindex $app$: given by the approximation Eq.  (\ref{fdip11}).  $t$ in $fs$. \bb}
\end{figure}

\begin{figure} [htbp]
\centerline{\includegraphics[width=1.0\columnwidth]{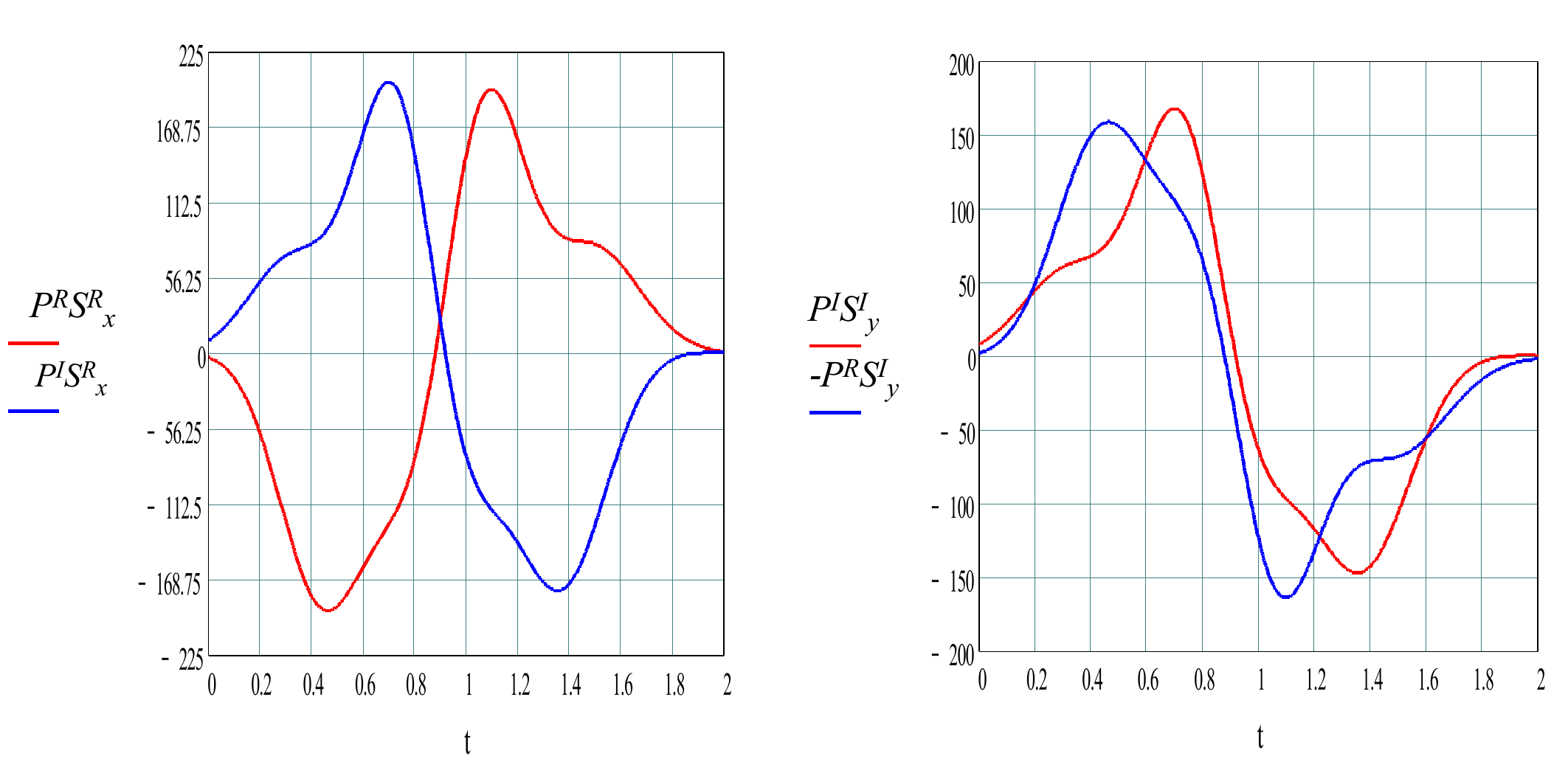}}
\caption{Ag particle with radius $a=5nm$. $x$ and $y$-components of the active and reactive forces ${\bf F}^R({\bf r}, t)$ and ${\bf F}^I({\bf r}, t)$, respectively,  (in $ pN$), due to the Poynting momentum, [cf. last two terms of Eqs. (\ref{fdip10_1})-(\ref{fdip10_2})], ${\bf r}={\bf  0}$.  Left:  $P^R S_x^R=\frac{1}{2c}\Re[\partial_t({\mathcal P}^*(t)\mathcal{E}(t))]\Re[{\bf E}_0({\bf r})\times {\bf B}_0^*({\bf r})]$ (active) and $P^IS_x^R=\frac{1}{2c}\Im[\partial_t({\mathcal P}^*(t)\mathcal{E}(t))]\Re[{\bf E}
_0({\bf r})\times {\bf B}_0^*({\bf r})]$ (reactive).  Right: $P^I S_y^I=\frac{1}{2c}\Im[\partial_t({\mathcal P}^*(t) \mathcal{E}(t))]\Im[{\bf E}_0({\bf r})\times {\bf B}_0^*({\bf r})]$ (active) and $-P^R S_{y}^{I}=
-\frac{1}{2c}\Re[\partial_t({\mathcal P}^*(t)\mathcal{E}(t))]\Im[{\bf E}_0({\bf r})\times {\bf B}_0^*({\bf r})]$ (reactive).  $t$ in $fs$.}
 \end{figure}

\begin{figure} [htbp]
\centerline{\includegraphics[width=1.0\columnwidth]{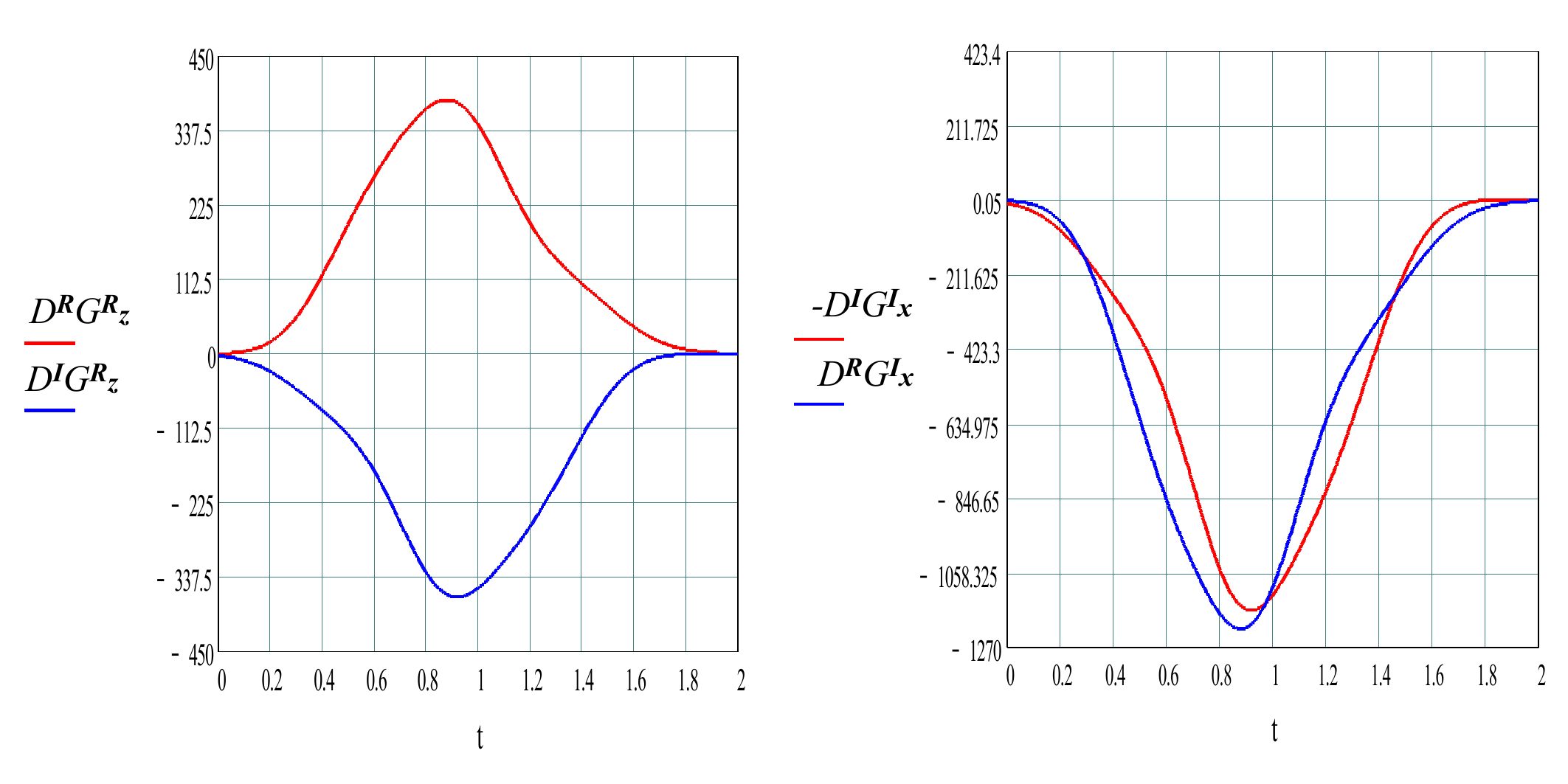}}
\caption{Ag particle with radius $a=5nm$. $x$ and $z$-components of the active and reactive forces ${\bf F}^R({\bf r}, t)$ and ${\bf F}^I({\bf r}, t)$, respectively,  (in $ pN$), due to the the gradient parts, [cf. first two terms of Eqs. (\ref{fdip10_1})-(\ref{fdip10_2})],  ${\bf r}={\bf  0}$.  Left:  $D^R G_z^R=\frac{1}{2}\Re[{\mathcal P}^*(t)\mathcal{E}(t)]\Re[{ E}_{0\, j}^*({\bf r})\partial_i {E}_{0\, j}({\bf r})]$ (active) and $D^I G_z^R=\frac{1}{2}\Im[{\mathcal P}^*(t)\mathcal{E}(t)]\Re[{ E}_{0\, j}^*({\bf r})\partial_i { E}_{0\, j}({\bf r})]$  (reactive).  Right:  $-D^I G_x^I=-\frac{1}{2}\Im[{\mathcal P}^*(t)\mathcal{E}(t)]\Im[{ E}_{0\, j}^*({\bf r})\partial_i { E}_{0\, j}({\bf r})]$ (active) and  $D^R G_x^I=\frac{1}{2}\{\Re[{\mathcal P}^*(t)\mathcal{E}(t)]\Im[{E}_{0\, j}^*({\bf r})\partial_i { E}_{0\, j}({\bf r})]$ (reactive). $t$ in $fs$.  }
\end{figure}

\begin{figure} [htbp]
\centerline{\includegraphics[width=1.0\columnwidth]{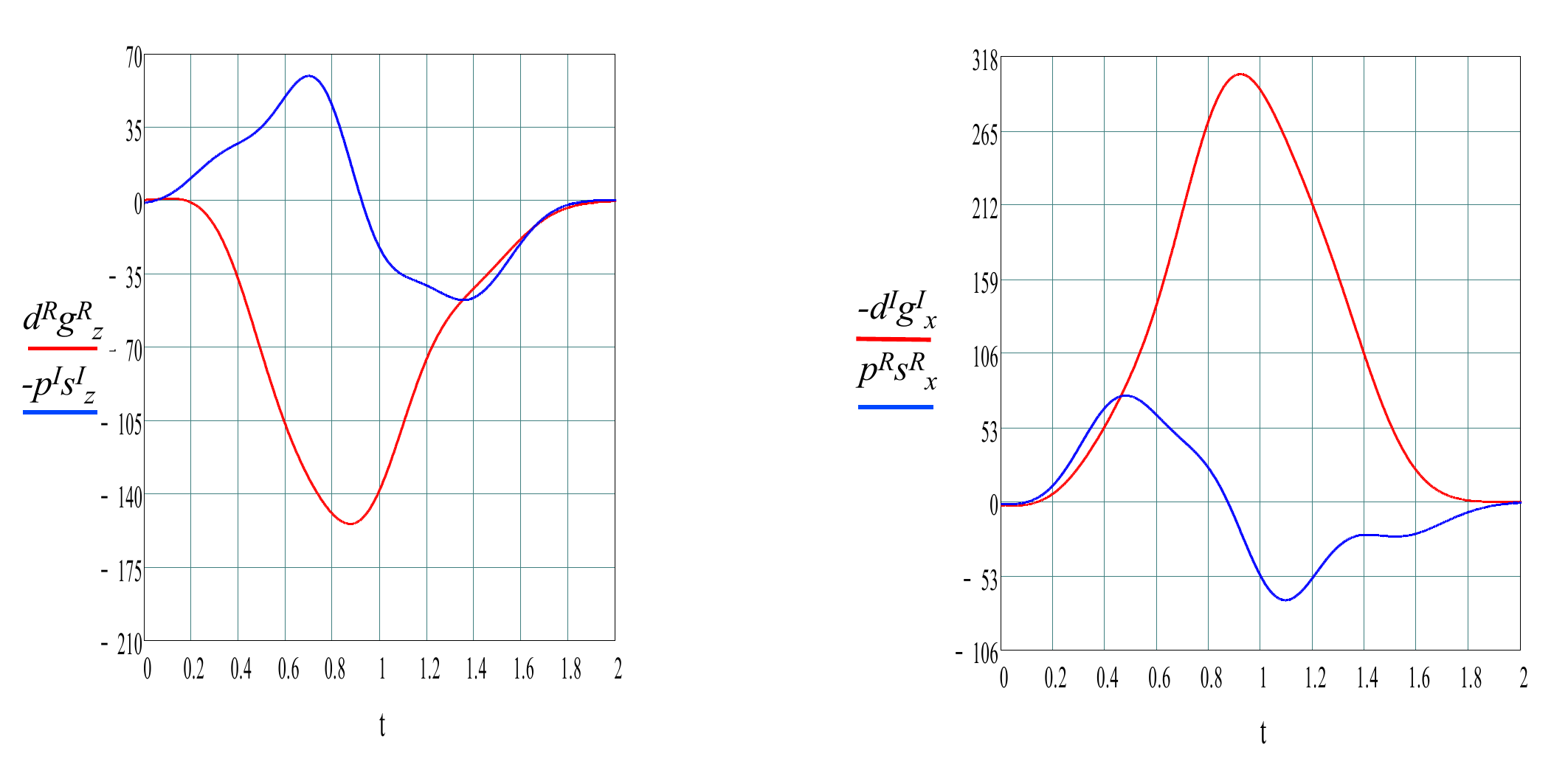}}
\caption{ Ag particle with radius $a=5nm$. Cartesian components of  ${\bf F'}({\bf r}, t)$, [cf. Eq. (\ref{fdip10_1prim})], (in $ pN$),  ${\bf r}={\bf  0}$.  Left:
 $d^R g_z^R=\frac{1}{2}\{\Re[{\mathcal P}(t)\mathcal{E}(t)]\Re[{ E}_{0\, j}({\bf r})\partial_i {E}_{0\, j}({\bf r})]$, $-p^I s_z^I= -\frac{1}{2c}\Im[\partial_t({\mathcal P}(t)\mathcal{E}(t))]\Im[{\bf E}_0({\bf r})\times {\bf B}_0({\bf r})]_i$. Right: $-d^I g_z^I=-\frac{1}{2}\Im[{\mathcal P}(t)\mathcal{E}(t)]\Im[{ E}_{0\, j}({\bf r})\partial_i { E}_{0\, j}({\bf r})]$,
$-p^R s_x^R= -\frac{1}{2c}\Re[\partial_t({\mathcal P}(t)\mathcal{E}(t))]\Re[{\bf E}_0({\bf r})\times {\bf B}_0({\bf r})]_i$
  $t$ in $fs$. \bb}
\end{figure}

\begin{figure} [htbp]
\centerline{\includegraphics[width=1.0\columnwidth]{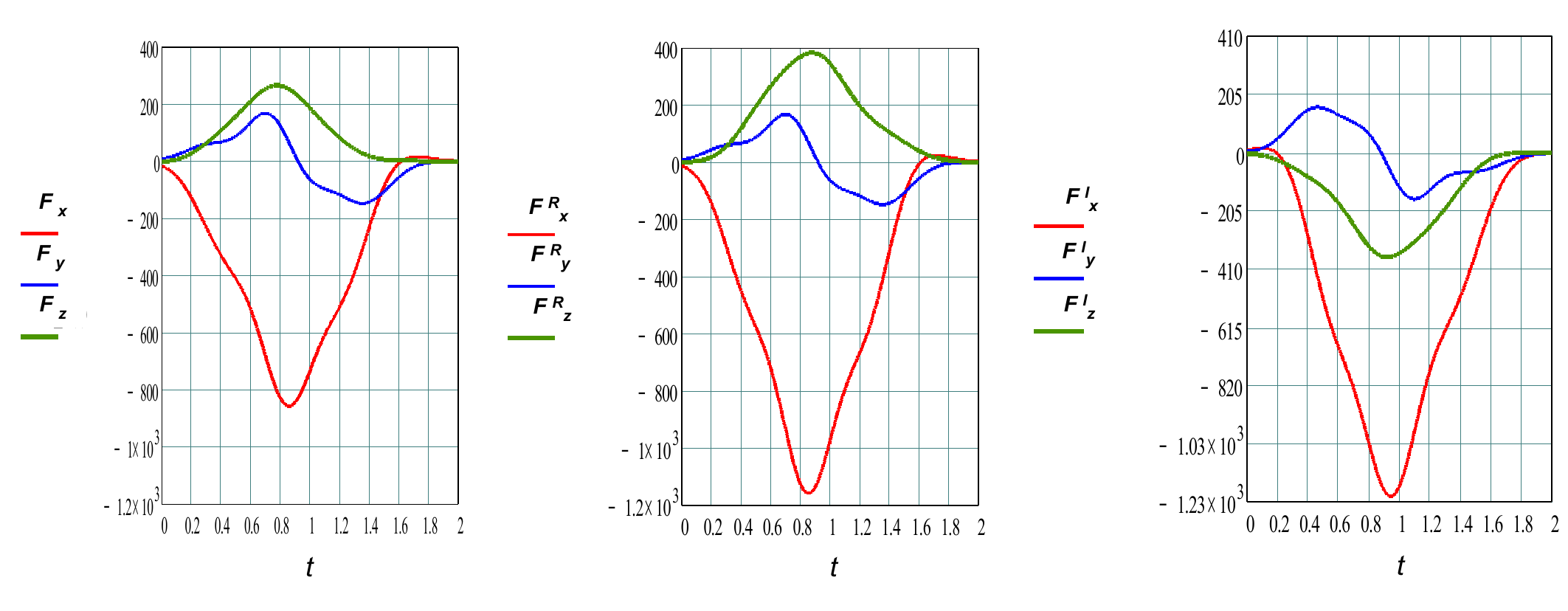}}
\caption{ Ag particle with radius $a=5nm$.  Cartesian components of the resultant instantaneous force, ${\bf F}({\bf r}, t)$,  (left), active force,
${\bf F}^R({\bf r}, t)$, (middle), and  reactive force, ${\bf F}^I({\bf r}, t)$, (right);  
  (in $ pN$),    $t$ in $fs$.  ${\bf r}={\bf  0}$. \bb}
\end{figure}

\begin{figure} [htbp]
\centerline{\includegraphics[width=1.0\columnwidth]{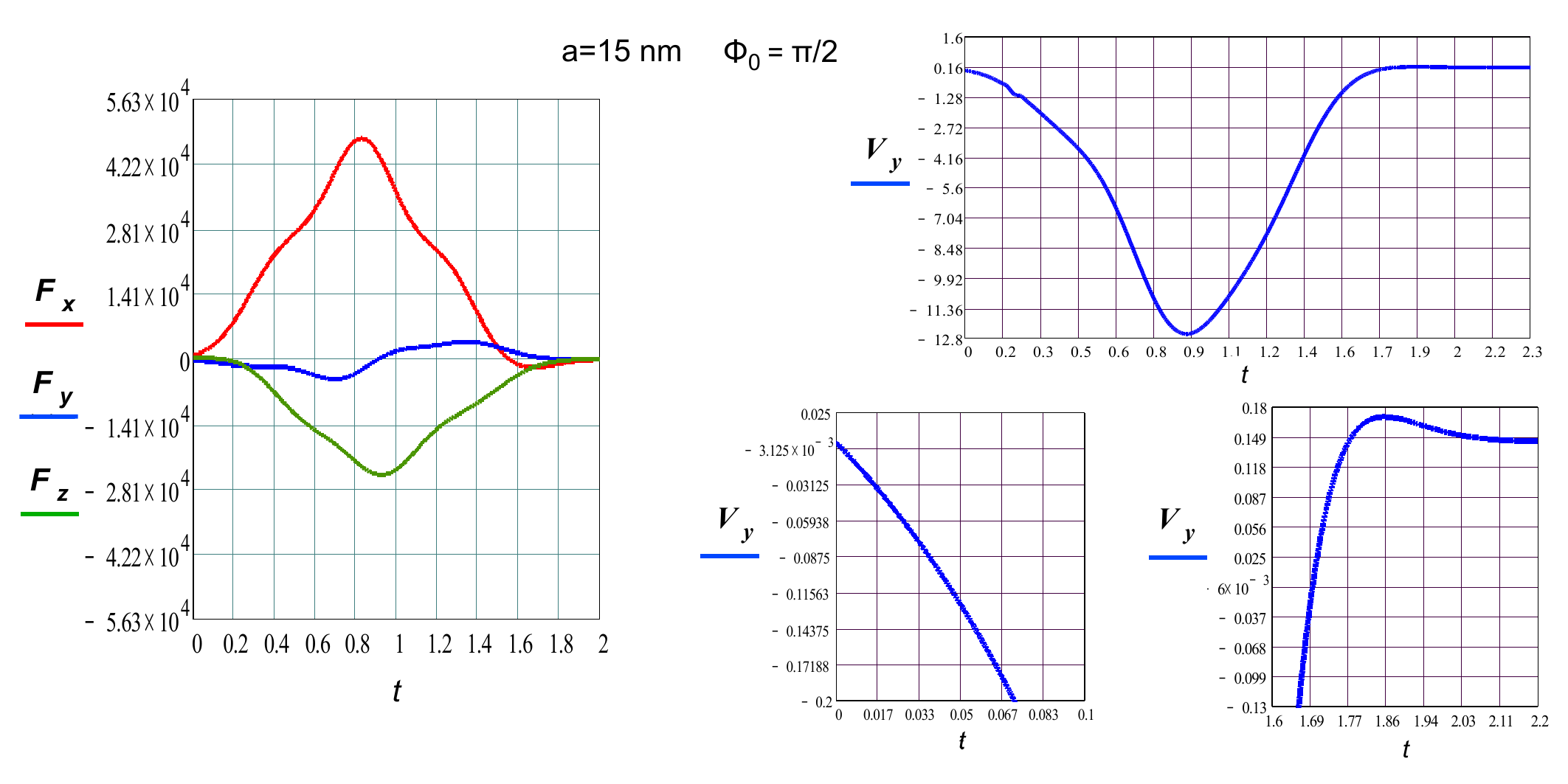}}
\caption{ Ag particle with radius $a=15nm$. Left: Cartesian components of the resultant instantaneous force, ${\bf F}({\bf r}, t)$,    (in $ pN$),    $t$ in $fs$.  ${\bf r}={\bf  0}$. Upper right: Lateral velocity  ${\cal  V}_y(t)$ of the particle (in $ \mu m/s$), as the pulse evolves in time.   $t$ in $fs$. Lower right: details of the evolution of ${\cal  V}_y(t)$ at the beginning and at the end of the  evanescent wave pulse. \bb}
\end{figure}

\subsection{Time-resolved forces from an  subcycle pulsed evanescent wave.}
We  consider a pulse with $\tau<T_0=1/\nu_0$  in $z<0$ that  by total internal reflection (TIR)  at a plane interface $z=0$ separating  air ($\epsilon=\mu=1$)  in $z>0$ from a dielectric in the half-space $z\leq 0$, generates an evanescent pulse  of the form (\ref{fdip3bis}) in  $z>0$.   The plane of incidence is $OXZ$. Then  the complex spatial parts of the electric and magnetic vectors in $z>0$, are expressed in a Cartesian coordinate basis $\{\hat{\bf x},\hat{\bf y},\hat{\bf z}\}$ as 
 \cite{cpv}:
\be
{\bf E}_0({\bf r})= \left(-\frac{iq}{k } T_{ \parallel}, T_{\perp},
\frac{K}{k } T_{\parallel}\right) \exp(iKx-qz), \,\,\, \nonumber \\
{\bf B}_0({\bf r})=  \left(-\frac{iq}{k } T_{ \perp}, -
T_{\parallel}, \frac{K}{k } T_{ \perp}\right)  \exp(iKx-qz) .\,\,\,\, \label{evan}
\ee 
For TE or $s$ (TM or $p$) - polarization , i.e. ${\bf E}_0$ (${\bf B}_0$) perpendicular to the plane of incidence $OXZ$, only those components with the (non-dimensional)  transmission coefficient $T_{ \perp}$, ($T_{\parallel}$) would be chosen in the incident fields \cite{cpv}. $ K$  denotes the component, parallel to the interface, of the wavevector ${\bf k}$: $k({\bf s}_{xy}, s_z)= ( K,
0, iq)$, $q= \sqrt{K^2 -k ^2}$, $k ^2= K^2-q^2$.

Now, it is known \cite{cpv,bliokh1} that being $w$, $w_{react}$, $\hel$ and $\qui $ the energy, reactive power, helicity and reactive helicity \cite{cpv,helitorque} of the wave, respectively,  the complex Poynting vector of this evanescent wave  is:
\be
{\bf S}=\frac{c}{8\pi\mu}[\frac{K}{k}(| T_{ \perp}|^2+|T_{\parallel}|^2)\,, i 2\frac{Kq}{k^2}T_{\perp}^*T_{\parallel}\, , -
 i \frac{q}{k}(| T_{ \perp}|^2-|T_{\parallel}|^2)] 
\exp(-2qz) \,\,\,\,\,\,\,\,\,\,\nonumber \\
=[\frac{ck}{ K} w\,, \frac{c}{4\pi }(-\frac{kq}{K}\hel + 
 i \frac{kK}{q}\qui)  , 
-\frac{i}{2 \mu  q}w_{react}];
 \,\,(\mu=1).\,\,\,\,\,\,\,\,\,\,\,\label{CPevan}
\ee
And hence it has a transversal $y$-component associated to $\Im[{\bf E}_0({\bf r})\times {\bf B}_0^*({\bf r})]$ that through the fourth term of  (\ref{fdip10_1}) and third term of (\ref{fdip10_2}) will give rise to  an instantaneous first-order transversal force on a purely electric dipolar particle,  (and should analogously  be observed on a purely magnetic dipolar particle \cite{cpv,bliokh1}).  Thus, as remarked above,  we next study the temporal factors  $(1/2)[{\mathcal P}^*(t)\mathcal{E}(t)]$ and $(1/2c)[\partial_t({\mathcal P}^*(t)\mathcal{E}(t))]$ that modulate these instantaneous forces.

We  also  study the effect of the time envelope in the $x$ and $y$-component of the force from the complex Poynting momentum in the last two terms of the instantaneous forces ${\bf F}^R({\bf r}, t)$ and ${\bf F}^I({\bf r}, t)$, Eqs. (\ref{fdip10_1})-(\ref{fdip10_2}). TIR of a plane wave, incident at $60^{\circ}$  on a  fused silica ($n=1.5$)$/$air plane interface, with diagonal linear polarization at $45^{\circ}$.  The corresponding evanescent wave at $x=z=0$ has $K=1.3k$ and $q=0.83k$. $T_{\parallel}=T_\perp$, $T_{\perp}=k/(\sqrt{2} K)$. So that $\frac{1}{2c}\Re[{\bf E}_0({\bf r})\times {\bf B}_0^*({\bf r})]=\frac{1}{2c}(\frac{k}{K},0,0)$,  $\frac{1}{2c}\Im[{\bf E}_0({\bf r})\times {\bf B}_0^*({\bf r})]=\frac{1}{2c}(0,\frac{q}{K},0)$, which shows that the  $OX$-component of the Poynting momentum is real while its $OY$-transversal component  is imaginary \cite{cpv, bliokh1}. Moreover  we have $||{\bf E}_0 \times {\bf B}_0^*||=1$. Therefore
the instantaneous intensity of this evanescent pulse, incident on the particle, is
\be
I_0=|{\bf S}_0|=\frac{c}{4\pi} |\mathcal{E}(t-\frac{x}{c})|^2 |{\bf E}_0 \times {\bf B}_0^*|=\frac{c}{4\pi} |\mathcal{E}(t-\frac{x}{c})|^2. \label{instantI}
\ee
The resulting time-varying forces depend on the c-phase $\phi_0$.  In what follows we shall consider the particle location at  ${\bf r}= {\bf 0}$ and $\phi_0=0$ to get a qualitative picture of the  temporal factors that govern these time-varying forces. Let this evanescent pulse have a carrier c-wavelength $\lambda_0=438nm$ , equivalent to $2.83 eV$ \cite{halaspage}, ($\nu_0=685 THz$, $\omega_0=2\pi\nu_0=4304 THz$, $T_0=\lambda_0/c=1460 as$, $c=299.79 nm/fs$) and width (full width at half maximum, FWHM)  $\tau=365 as=0.25T_0$, being incident on an Ag sphere in the air, resting  with its center at $x=0$ on the TIR interface $z=0$.  We shall initially address two situations corresponding to particle radius: $a=5 nm$ and $a=20 nm$.  Because of the high c-frequency of the pulse,  we have assumed such TIR evanescent wave being  created on a fused quartz prism whose refractive index at this deep UV wavelength is $1.51$.

We take $A_0=10$, which implies assuming the pulse {\it  peak power}: $Max(I_0)=1.63KW/\mu m^2$  at the focus of a Gaussian beam  with spot size  of radius  $R_0=50$ $\mu m$  on the air/silica interface. Then, according to (\ref{fdip4bis}) and  (\ref{instantI}),   $Max(I_0) $ yields $Max(\mathcal{E}^2(t))=(8.26)^2\times10^{-12} J/ \mu m^3$. I.e,  $Max(\mathcal{E}(t))=8.26\times10^{-9}\times10^{-3/2} J^{1/2}/ nm^{3/2}$.

The {\it pulse energy} being  $E_p=1.7nJ$, so that \cite{chang,chang1} $Max(I_0)=1.88 E_p/(\tau \pi R_0^2)$, and the beam pulse {\it averaged power} is $I_{ave}=f_r E_p$, $I_{ave}=17mW$ with a {\it repetition rate } $f_r=10MHz$.  Figure 1 depicts  $\Re [\mathcal {E}_0(t)]$, and  envelope of this  pulse, $\mathcal {E}(t)$, ($x=0$, $\phi_0=0$), as well as the  c-wave: $\mathcal {C} (t)=A_0\Re[\exp(-i\omega_0 t)]$.

The temporal envelope $ {\mathcal P}^*(t)$, Eq. (\ref{fdip10}), of the induced electric dipole moment may be directly obtained by the inverse Fourier transform of $ {\mathcal P}^*(\nu)$, Eq. (\ref{fdip9}). We shall compare its exact variation with $t$, so obtained, with an approximation to $ {\mathcal P}^*(\omega)$, (\ref{fdip9}), around   the resonant frequency $\omega_r=2\pi \nu_r$, and hence  we make the usual expansion   \cite{landau1,milonni1}:  $\mathcal {E^*}(\nu)=\mathcal {E^*}(\nu_r)+\partial_{\nu}\mathcal {E^*}(\nu_r)(\nu-\nu_r)$. With this approximation of $\mathcal {E^*}(\nu)$, via $\int_{-\infty}^{\infty} d\nu\exp(i2\pi\nu t)\mathcal {E^*}(\nu)\alpha^*(\nu)$,  writing $\alpha^*(\nu)$ as $\alpha^*(\nu-\nu_r)=\alpha^*(\nu)*\delta(\nu-\nu_r)$ in order to explicitely show  the oscillation of ${\mathcal P}^*(t)$ with period $1/\nu_r$, (the polarizability resonance being at  $\nu=\nu_r$) , we get 
\be
{\mathcal P}^*(t)\simeq \exp(2\pi i \nu_r t)[\mathcal {E^*}(\nu_r){\alpha}^*(t)-\frac{i}{2\pi}\partial_{\nu}\mathcal {E^*}(\nu_r)\partial_t\alpha^{*}(t)]. \label{fdip11}
\ee

Fig. 2 illustrates the time variation  of the real and imaginary parts of ${\mathcal P}^*(t)$, both with  its exact values from direct  Fourier inversion of $\mathcal{P}^*(\nu)=\mathcal{E}^*(\nu)\alpha^*(\nu)$,  Eq.(\ref{fdip9}), and with subindex $app$ given by its approximation (\ref{fdip11}). Two cases of sphere radius: $a=5nm$ and $a=20nm$ are addressed \cite{footnote}. 
As seen, at very small values of $t$, ($t<0.2 fs$), there is a failure of the approximated dipole envelope, which is a consequence of  the departure of the approximation (\ref{fdip11}) from its exact values at higher frequencies:  $|\nu-\nu_r|>3$ where obviously the  small parameter Taylor expansion fails no matter one took higher order terms.  Notice the envelope oscillation of  $\mathcal {P}^*(t)$, in agreement with the factor $\exp(2\pi i \nu_r t)$ due to the effect in the inverse Fourier transform of the shift of $\nu_r$ with  respect to $\nu=0$ of  (\ref{fdip11}), whose period is: $1/\nu_{r}=0.8 fs$ for both spheres since they have  resonant frequency close to $\nu_{r}=1260 THz$. 

  In this connection, we remark that although the c-frequency $\nu_0$ is far from  $\nu_{r}$, the $\omega^2\tau^2$ factor in Eq. (\ref{fdip5bis}) enhances the amplitude maximum of $\mathcal {E}(\nu)$ and shifts it near  $\nu_{r}$. Thus   the Ag sphere resonance is excited. \bb

The sphere radius  influences the magnitude of  ${\mathcal P}(t)$,  which in turn will also affect the instantaneous forces; but it does slightly in the shape of this factor as $t$ increases. This is due to the shape of $\alpha^*(\nu)$ versus that of $ \mathcal{E}^*(\nu) $ and thus of their corresponding Fourier transforms. 
Fig. 3 shows the  real and imaginary parts of the temporal  envelope $(1/2){\mathcal P}^*(t)\mathcal{E}(t)$, which appears in the first two  terms of  (\ref{fdip10_1})-(\ref{fdip10_2}) and that modulate the gradient and the orbital momentum forces \cite{LSA22, cpv, nieto1,bliokh1}. There we observe departures of the approximated analytic expression (\ref{fdip11}) for $t<0.19 fs$. In addition,  there is an  oscillation of $\Im[{\mathcal P}^*(t)\mathcal{E}(t)]$ changing  sign  between $0.2 fs$ and  $0.5 fs$ where  it becomes negative; thus  giving rise to a {\it repulsive time-resolved gradient-component of}  ${\bf F}^I({\bf 0},t)$, and to a {\it pulling scattering orbital momentum-component of} ${\bf F}^R({\bf 0},t)$, in contrast with  the time-averaged  force from a monochromatic Gaussian evanescent wave  \cite{LSA22,cpv, bliokh1}.

The change of sign of the time modulation factor has even more dramatic effects in the last two terms of  (\ref{fdip10_1})-(\ref{fdip10_2}), as depicted in  Fig. 4, where the factor $(1/2c)\partial_t({\mathcal P}^*(t)\mathcal{E}(t))$ is shown. This envelope, whose shape is seen (on comparison with Fig.1)  to be dominated by that of the pulse, Eq.(\ref{fdip4bis}), modulates in time the contribution of both the real and imaginary parts of the complex Poynting momentum ${\bf g}({\bf r})=(1/c^2){\bf S}({\bf r})$. There it is seen  the interesting feature where the imaginary part of this quantity is minimum and negative while the real part is small and positive, and viceversa. Thus having a dramatic effect in the sign of this component of the  force.

 We observe in these figures a gradual departure of the approximation Eq. (\ref{fdip11}) as the size of the particle increases. But this size cannot increase indefinitely within the range of validity of the dipole approximation.  In fact, while the sphere of $a=5 nm$ is a Rayleigh one since $ka=0.132$, the one with $a=20 nm$ holds $ka=0.528$ which barely fulfills the Rayleigh criterion $ka<<1$; and thus the calculations for this particle are done with its electric polarizability $\alpha(\omega)$ in terms of the first electric Mie coefficient $a(\omega)$, (see Eq. (34) of \cite{nieto1}). Namely, $\alpha(\omega)=  i(3/2k^3)a(\omega)$. 

 As a limiting case within correctly assuming the particle as dipolar, Fig. 5 illustrates a large departure of the approximation Eq.(\ref{fdip11}) in both  $(1/2){\mathcal P}^*(t)\mathcal{E}(t)$ and $(1/2c)\partial_t({\mathcal P}^*(t)\mathcal{E}(t))$  for an Ag sphere with $a=50 nm$. \bb

Returning to the case of the Ag sphere with radius  $a=5nm$ and incident   wavelength $\lambda_0=238 nm$,  we show in Fig. 6 both the exact and approximated  real and imaginary parts of the terms  $(1/2){\mathcal P}(t)\mathcal{E}(t)$ and  $(1/2c)\partial_t({\mathcal P}(t)\mathcal{E}(t))$ of   ${\bf F}'$, Eq.(\ref{fdip10_1prim}).

Figure 7  illustrates the effect of the time envelope, $\partial_t({\mathcal P}^*(t)\mathcal{E}(t))$, on the $x$-component (associated to $\Re[{\bf E}_0({\bf r})\times {\bf B}_0^*({\bf r})]$) and $y$-component  (associated to $\Im[{\bf E}_0({\bf r})\times {\bf B}_0^*({\bf r})]$) of the  active and reactive  forces ${\bf F}^R({\bf r}, t)$ and ${\bf F}^I({\bf r}, t)$, Eqs. (\ref{fdip10_1})-(\ref{fdip10_2}).
 $P^R S_x^R=\frac{1}{2c}\Re[\partial_t({\mathcal P}^*(t)\mathcal{E}(t))]\Re[{\bf E}_0({\bf r})\times {\bf B}_0^*({\bf r})]$, -$P^R S_{y}^{I}=
-\frac{1}{2c}\Re[\partial_t({\mathcal P}^*(t)\mathcal{E}(t))]\Im[{\bf E}_0({\bf r})\times 
{\bf B}_0^*({\bf r})]$,  $P^I S_y^I=\frac{1}{2c}\Im[\partial_t({\mathcal P}^*(t)
\mathcal{E}(t))]\Im[{\bf E}_0({\bf r})\times {\bf B}_0^*({\bf r})]$ and 
$P^IS_x^R=\frac{1}{2c}\Im[\partial_t({\mathcal P}^*(t)\mathcal{E}(t))]\Re[{\bf E}
_0({\bf r})\times {\bf B}_0^*({\bf r})]$ are shown. 
The red lines correspond to  ${\bf F}^R({\bf r}, t)$, Eq. (\ref{fdip10_1}), while the blue lines are the components of  ${\bf F}^I({\bf r}, t)$, Eq. (\ref{fdip10_2}). It is interesting that in this contribution of the imaginary Poynting momentum along $OY$,  the shape of this component of  ${\bf F}^R$ is qualitatively similar to that of ${\bf F}^I$, although the later is time-shifted. However, along $OX$ where there is contribution of the  real momentum,  the shapes of the correspnding lateral components of  the forces ${\bf F}^R$ and  ${\bf F}^I$ from this pulsed evanescent wave, also being t-shifyed, oppose to each other.  

This lag-behind and/or opposition of  ${\bf F}^I$  upon ${\bf F}^R$  characterizes the reactive character of the ILF, and keeps an analogy in the area of nano-optics with the influence of the reactive energy, due to the flow of imaginary Poynting momentum, on the scattered or radiated cw-energy from nanoparticles or nano-antennas, expressed by the flow of real (time-averaged) Poynting momentum \cite{cpv,navarra}.

 The transversal $y$-component of the forces on the dipolar particle due to  $\Im[{\bf E}_0({\bf r})\times {\bf B}_0^*({\bf r})]$ is interesting since we see in Fig. 7 that  when time-resolved, it may acquire negative values, thus being opposite to the usual time-averaged transversal force observed in magnetodielectric particles impinged by a monochromatic evanescent wave at the same polarization \cite{cpv,bliokh1}.

An analogous behavior is observed in Figs. 8 and 9 with the $x$ and $z$-components due to the gradient and orbital momentum of ${\bf F}^R({\bf r}, t)$,  ${\bf F}^I({\bf r}, t)$ and ${\bf F}'({\bf r}, t)$. It is interesting that while  the  $z$-component   $D^IG_z^R$ of  ${\bf }F^I({\bf r}, t)$ may be attractive,  $-p^I s_z^I$  and $p^R s_z^R$ of  ${\bf F'}({\bf r}, t)$,  may  change sign. The latter enter in the instantaneous force, [cf. Eqs. (\ref{fdip10_1r}) and  (\ref{fdip10_1prim})].

 Figure 10 exhibits the resultant forces: active, ${\bf F}^R({\bf 0}, t)$,  reactive, ${\bf F}^I({\bf 0}, t)$, and instantaneous, ${\bf F}({\bf 0}, t)$, on the Ag sphere of $a=5nm$ at rest in ${\bf r}={\bf  0}$. A repulsive  $z$-instantaneous force is observed at any $t$, although the sign of this component   may be controlled with the structure,  $t$-dependence of the pulse  and  its carrier phase $\phi_0$,  as we shall next see. Some reactive Cartesian components may either oppose  those of both the active and instantaneous forces, or be $t$-shifted. The attractive, or repulsive (depending on the range of $t$) lateral $y$- component of  the instantaneous force, which is fully due to the $y$-component of the active force, is also remarkable. Also, we observe an intriguing pulling  $x$-component of these active and instantaneous forces, which opposes to the direction of the Poynting and canonical momenta of the evanescent wavefield as a consequence of its t-dependence. In addition, there is a remarkable levitating nature of both the instantaneous and active $z$-components. 

\section{optical transportation. Lateral movement,  pulling and levitation of an electric dipole particle}
The lateral $y$-component, and the pulling force along $OX$ of the instantaneous ${\bf F}({\bf 0}, t)$  from the pulsed evanescent wave, makes it possible   to deliver both a lateral and pulling dynamics on electric dipole particles. So far this was only known to be possible in evanescent waves acting on magnetoelectric particles, i.e. with both electric and magnetic induced dipoles \cite{cpv, bliokh1}, but not on purely electric ones. 

Concerning the former, let us look at the last two terms of Eqs. (\ref{fdip10_1}) and (\ref{fdip10_1prim}). We can consider  the force averaged over the duration, $\sigma$  of the pulse, (in contrast with the usual period-averaging in monochromatic fields), which should be an observable. Namely:
\be
\frac{1}{\sigma}\int_0^\sigma dt{\bf F}({\bf 0},t)=\frac{1}{2\sigma}\int_0^\sigma dt\{\Re[{\mathcal P}^*(t)\mathcal{E}(t)]\Re[{ E}^*_{0\, j}({\bf r})\partial_i {E}_{0\, j}({\bf r})]-\Im[{\mathcal P}^*(t)\mathcal{E}(t)]\Im[{ E}^*_{0\, j}({\bf r})\partial_i { E}_{0\, j}({\bf r})] \nonumber \\
+\Re[{\mathcal P}(t)\mathcal{E}(t)]\Re[{ E}_{0\, j}({\bf r})\partial_i {E}_{0\, j}({\bf r})]-\Im[{\mathcal P}(t)\mathcal{E}(t)]\Im[{ E}_{0\, j}({\bf r})\partial_i { E}_{0\, j}({\bf r})] \nonumber \\
+\frac{1}{c}\{\Re[({\mathcal P}^*(t)]\mathcal{E}(t))]_{\sigma}-\Re[({\mathcal P}^*(t)]\mathcal{E}^*(t))]_{0}\}\Re[{\bf E}_0({\bf r})\times {\bf B}^*_0({\bf r})]_i -\frac{1}{c}\{\Im[{\mathcal P}^*(t)\mathcal{E}(t))]_{\sigma}-\Im[{\mathcal P}^*(t)\mathcal{E}(t))]_{0}\}\Im[{\bf E}_0({\bf r})\times {\bf B}^*_0({\bf r})]_i\}\nonumber \\
 \nonumber \\
+\frac{1}{c}\{\Re[({\mathcal P}(t)]\mathcal{E}(t))]_{\sigma}-\Re[({\mathcal P}(t)]\mathcal{E}(t))]_{0}\}\Re[{\bf E}_0({\bf r})\times {\bf B}_0({\bf r})]_i -\frac{1}{c}\{\Im[{\mathcal P}(t)\mathcal{E}(t))]_{\sigma}-\Im[{\mathcal P}(t)\mathcal{E}(t))]_{0}\}\Im[{\bf E}_0({\bf r})\times {\bf B}_0({\bf r})]_i\}\}. \,\,\,\,\,\,\, \,\,\,\,\,\,\, \label{midforc}
\ee
We see that the last two terms in (\ref{fdip10_1}) and (\ref{fdip10_1prim}) do not contribute to this  averaged force (\ref{midforc}). This indicates no net momentum transferred to the particle from  the  Poynting momenta after the pulse goes away. 

Notwithstanding, this does not imply the particle longitudinal and lateral displacements $\Delta x$ and $\Delta y$, respectively,  due to  these terms are zero. For example assume the Ag dipolar particle to be initially at rest in vacuum on the TIR interface.  Looking at the transverse action of the four terms of (\ref{midforc}), during the pulse duration,  their effect will be first to accelerate the particle,  even changing its direction of $y$-movement, then decelerating it; so that finally it acquires a constant velocity. An analogous effect will arise in the $x$-direction; this time constantly pulling during the duration of the pulse, as shown in Fig. 10 for the $5nm$ particle. Thus by this process, the accumulated displacement is nonzero. On the other hand, the pushing effect of the $F_z^R$ and $F_z$ components observed in Fig. 10, indicates that it levitates the particle away from the TIR interface when it is subjected to the constant repetition of the pulsed evanescent wave; which is a novel phenomenon in contrast with the well-known attractive  gradient $z$-force from evanescent continuous waves.

However, it is straightforward to see that the sign of these forces can easily be controlled by e.g. choosing the shape of the pulse through the shift $x/c$ and carrier phase $\phi_0$. For example, Fig. 11 illustrates the instantaneous force Cartesian components on an Ag sphere of radius $a=15 nm$ at rest on the TIR interface , illuminated by the same wavefield as before but now with the choice $x/c=0.9$ and $\phi_0=\pi/2$. We observe forces two orders of magnitude larger than those shown in Fig. 10 for the $5nm$ particle; and with their sign inverted with respect to those of Fig. 10.  

Let us look for example at the   lateral velocity ${\cal  V}_y(t)=(1/M)\int dt F_y(t)$ acquired by the $15 nm$ Ag particle along $OY$ during the duration $\sigma$ of the pulse. The mass $M$ of this  sphere has been estimated to be $M= (4\rho/3)\pi a^3=1.49 \times 10^{-16}$ grams, ($a=15 nm$, $\rho=10.5 g/cm^3$), After the pulse finishes at about $t=200 As$, the Ag particle remains at a constant velocity ${\cal  V}_y=+0.15 \mu m/s$, as shown in Fig. 11. Since  the pulse repeats once again every $10^{-7} s$, it is an easy exercise to see that its accumulated multiple kicks  wil lead to a much larger velocity  ${\cal  V}_y$, along with an observable net lateral displacement $y(t)=\int dt {\cal  V}_y(t)$ of the particle after one second has elapsed.

A similar analysis will show an important net displacement of the particle along the directions $OX$ of propagation of the pulsed wave, and along $OZ$, normal to the TIR interface. Figures 10 and 11 suggest  that the velocity   ${\cal  V}_x$ of the sphere will  point either against or  along the canonical momentum, respectively. The negative  ${\cal  V}_x$ implied by the pulling instantaneous, ${\bf F}_x$, and active,  ${\bf F}_x^R$ in Fig. 10, is intriguing. The same may be said of the dynamics of the particle along $OZ$, which while Fig. 11 shows an attraction towards the interface, Fig. 10 suggests a levitating $z$ force.

\section{Conclusion}
Time-dependent electrodynamical forces from subcycle ultrafast pulses belongs to an area where previous studies on over-cycle pulses cannot be applied. Also it highlights the novel appearence of two forces in the manipulation of nanostructures in the attosecond regime. Although we have illustrated them with their action on a resonant Ag sphere, so that they are enhanced, other particles, either in resonant or non-resonant regimes of the illuminating field wavelength, may be dealt with.  The variation with time of the reactive force shows how it affects both the active and instantaneous forces. However, given the so far unknown area occupied by the active and reactive forces beyond the recent study of \cite{LSA22},  we are only able to  interpret them as generalizations of the time-average and imaginary Lorentz force, respectively,  from time-harmonic fields to the realm of  time-dependent electromagnetic fields. More theoretical and experimental research is necessary to fully  identify their respective influence and relevance in electrodynamical forces, and hence in the area of optical manipulation of matter. On the whole, this study uncovers the mechanical action of subcycle pulses.

 The novel consequence of this study, which demostrates the versatility of attosecond pulses whose representation allows a factorization of their space and time-dependent  parts, leads to the appearence of   lateral, levitating, and pulling forces from evanescent waves, never observed before as far as we know; which is of  both fundamental and practical interest.  For one thing, the flexible  design of these pulses shows  the delivery of isolated electric dipoles, which was so far thought impossible, as well as the control of their directions of   movement. For another, it offers an  ultrafast route to  the precision of optical manipulation, where $nx$, $ny$ and $nz$ give the accuracy and  represent the translation distance; $n= 10^7$ being the number of pulses per second stricking the particle. 

We expect that this study may open the door of a new scenary in  ultrafast electrodynamics. Further research  should increase the insight and operation of the active and reactive forces put forward here, and their contribution into the unusual behaviour of instantaneous forces shown here. This is of importance to get  access to  the time-resolved dynamics and manipulation of nanostructures.
\bb

\setcounter{secnumdepth}{0}
\section{Acknowledgments}
We acknowledge support from the  National Natural Science Foundation of China (12274181), National Key R\&D Program of China (2023YFF0613700) and Ministerio de Ciencia, Innovación y Universidades of Spain (Grant PID2022-137569NB-C41).



\begin{thebibliography}{99}
\bibitem{footnote} 
The silver sphere has a polarizability:  $\alpha(\omega)=\alpha^{(0)}/[1-2ki\alpha^{(0)}/3] $; where  $\alpha^{(0)}(\omega)=\alpha^{(0)\,R}+i \alpha^{(0)\,I}=a^3\frac{\epsilon(\omega)-1}{\epsilon(\omega) +2}$ is the quasistatic polarizability with $\epsilon(\omega)$ obtained from \cite{johnson}. For example, for the one of radius  $a=5 nm$ , (in which  the plasma frequency $\omega_p=9.02 eV$, the relaxation time being $\tau=31 fs$), the resonant frequency is at $\omega_r=7920 THz=5.21 eV$, or $ \nu_r=1260 THz$.
\end{thebibliography}
\end{document}